\documentclass{aa}
\usepackage{graphics}
\usepackage{astron}
\newcommand{\Teff}{$T_{\rm eff} \ $}
\begin{document}
\title{\\ Subgiants as probes of galactic chemical evolution
\thanks{Based on observations made at ESO, La Silla}$^{,}$
\thanks{Based on observations made at NOT, La Palma}$^{,}$
\thanks{Tables 1 and 2 are only available in electronic form at the
CDS via anonymous ftp to cdsarc.u-strasbg.fr (130.79.128.5) or via
http://cdsweb.u-strasbg.fr/cgi-bin/qcat?/A+A/???/???}}
\titlerunning{Subgiants as probes of galactic chemical evolution}
\author{Patrik Thor\'en, Bengt Edvardsson and Bengt Gustafsson}
\authorrunning{Thor\'en et al.}
\institute{Department of Astronomy and Space Physics, Uppsala Astronomical
           Observatory, Box 515, S-751\,20 Uppsala, Sweden}

\date{Received 12 March 2004 / Accepted 28 May 2004}
\offprints{Bengt Edvardsson \\ (Bengt.Edvardsson@astro.uu.se)}
\abstract{
Chemical abundances for 23 candidate subgiant stars have been derived with the
aim at exploring their usefulness for studies of galactic chemical evolution.
High-resolution spectra from ESO CAT-CES and NOT-SOFIN covered 16 different
spectral regions in the visible part of the spectrum.
Some 200 different atomic and molecular spectral lines have been used for
abundance analysis of $\sim$ 30 elemental species.
The wings of strong, pressure-broadened metal lines were used for
determination of stellar surface gravities, which have been compared with
gravities derived from {\sc Hipparcos} parallaxes and isochronic masses.
Stellar space velocities have been derived from {\sc Hipparcos} and Simbad data,
and ages and masses were derived with recent isochrones.
Only 12 of the stars turned out to be subgiants, i.e. on the ``horizontal''
part of the evolutionary track between the dwarf- and the giant stages.
The abundances derived for the subgiants correspond closely to those of
dwarf stars.
With the possible exceptions of lithium and carbon we find that subgiant stars
show no ``chemical'' traces of post-main-sequence evolution and that they are
therefore very useful targets for studies of galactic chemical evolution.
\keywords{stars:abundances -- stars:evolution -- galaxy:abundances --
galaxy:evolution}}
\maketitle

\section{Introduction}

Abundance patterns in stellar populations have proven to be a powerful means of
studying galactic chemical evolution and nucleosynthesis.
Studies like Edvardsson et al. (1993, EAGLNT)\nocite{eaglnt:93},
Feltzing \& Gustafsson (1998)\nocite{feltzing},
Chen et al. (2000)\nocite{chen00} and Fuhrmann (2004)\nocite{fuhrmann04}
show that with a large number of stars the history of the chemical elements
can be mapped to disclose essentials of the star formation history at different
locations in the Galaxy.

Even with the impressive amount of data presented in
the papers mentioned above only a tiny sphere with radius some 100\,pc
around the Solar system has been explored with the high accuracy
achievable from dwarf-star spectra.
Knowledge of stellar kinematics enable extension of this sphere, but only
regions within 2\,kpc of the solar galactocentric radius have been
systematically probed by these stars.
Compared to the dimensions of the Galaxy, this is certainly a small region.
The generalization of results gained from studying stars that currently
neighbour the Solar system will very likely be locally biased.

The volume is limited by the low luminosities of the dwarfs used in the surveys,
and by the limited view in certain directions through the galactic disk.
Subgiants are 1--2 magnitudes brighter than dwarfs, which opens up
larger volumes for study.
With the use of VLT-sized telescopes, subgiant targets increase the
observable radius from the Earth by more than ten times, and e.g. allow studies
of the thick galactic disk where it dominates the stellar number density.
Subgiant stars are, due to their relatively rapid evolution, much more
rare than dwarfs.
However, with a volume increased $>1000$ times there will be enough objects
available for detailed chemical-evolution studies.
The problem with reddening will become more important for distant subgiants.
Therefore, future investigations must use methods designed for estimating
stellar parameters regardless of reddening or independent determination of
reddening, unless dust-free windows in the disk can be used.
For important studies of halo and disk stars close to the Galactic pole
the reddening problem is reduced.

The subgiant stars, although recognised as a specific stellar group
by 1930 (Str\"omberg 1930\nocite{Stromberg}), were not understood in terms of
stellar evolution until the 1950ies.
At the Vatican Conference on Stellar Populations in 1957, it was first realised
that the use of these stars offered possibilities for dating
the Galactic disk (see references in the interesting review of
the history of subgiants by Sandage, Lubin \& VandenBerg 2003\nocite{SLV}).
The facts that the isochrones separate nicely
for different ages in the subgiant region, and that they run almost
horizontally in the $T_{\rm eff}-M_{\rm bol}$ diagram, make the stars
particularly suitable for dating purposes. This requires, however, a good
determination of $M_{\rm bol}$, i.e. the stellar parallaxes,
or of the surface gravity.

In this work, based on high resolution spectroscopy data, we explore the use
of subgiants for studies of galactic chemical evolution by investigating a
sample of nearby subgiants and comparing the resulting abundances with those of
more unevolved dwarf stars.
We will conclude that they may represent the best suited stellar group for
this kind of study.

In Sects.\,\ref{obsred}--\ref{surfgrav} the observations, data reductions and
spectrum analysis are described.
In Sects.\,\ref{agesandmasses} and \ref{dynamics} the ages, masses and
dynamical properties of the stars are derived, while
Sect.\,\ref{errors} discusses different sources of errors in the analysis.
Sect.\,\ref{abund3} presents the abundance results, and
finally Sects.\,\ref{discuss} and \ref{conclude} give a discussion of the
results and the conclusions of this work.


\section{Observations and reductions}
\label{obsred}

\begin{figure}[ht]
 \resizebox{\hsize}{!}{\includegraphics{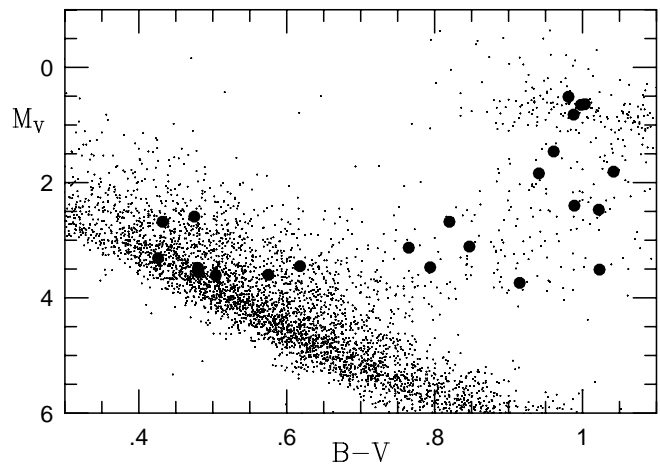}}
  \caption[]{The positions of our programme stars (large circles) in an
   observational HR diagram compared to {\sc Hipparcos} stars with relative
   parallax errors of 5\% or less}
 \label{hr_hipp}
\end{figure}

To find a test sample of {\it bona fide} subgiant stars
with a range of metallicities the compilation of [Fe/H] determinations
of Cayrel de Strobel et al. (1997) was used. Stars with luminosity
class IV and $\log g$ values around 3.0 (cgs) were found, and a number of
objects with different metallicities were chosen for possible observation.
At that time, the {\sc Hipparcos} (ESA, 1997) data was unfortunately not yet
available. After the release of these data several of our target stars turned
out not to be subgiants as shown by the colour-magnitude diagram in
Fig.\,\ref{hr_hipp}.
During the course of the analysis three stars were also found to be
Pop.\,II giants, HD\,4306, HD\,44007, and HD\,128279.
These stars were excluded from the discussion in the present paper.

The observational data were obtained during three observation runs in 1997.
The ESO CAT-CES at La Silla, Chile, was used in the periods March--April
(eight nights) and October--November (eight nights).
The NOT-SOFIN at La Palma was used in October--November (eight nights).
The observations at La Silla were made without problems.
Four nights of eight at NOT were useless due to bad weather and the great
'97 vegetation fire that lasted for one full night.

The spectra were reduced with the MIDAS software and routines.
The CES spectra consist of one echelle order only, and
were reduced with the MIDAS LONG context.
The SOFIN spectra have several orders.
These were extracted with the MIDAS ECHELLE context.

With the single-order CES the wavelength settings were changed several
times each night.
For every new setting during the observations a Th-Ar exposure was taken,
which was used for the wavelength calibration.

Scattered light background components were removed from the
exposures by subtracting a constant, where needed, for the CAT
spectra, and by the ECHELLE background fitting function
for the NOT multi-order echelle spectra.

\section{Stellar parameters}

Object selection and initial stellar parameter values were
based on the Cayrel de Strobel et al. (1997) compilation\nocite{cayrel97}.
which contains published data from hundreds of sources.
The stellar parameters had been derived by different
authors and some stars lacked a complete set of parameters.
Median values were used for objects with several parameter determinations.
During the autumn 1997 runs the {\sc Hipparcos} \cite{hipparcos} catalogue
had been released and subgiants could also be picked from there.
The latter catalogue does however not provide stellar parameters.
Literature search during the analysis provided scattered data
where, again, median values were used for reasonably modern
parameter determinations.
The atmospheric model parameters (except $T_{\rm eff}$) were modified
during the course of the analysis.
Surface gravities were determined from the wings of strong lines
(see Sect.\,\ref{surfgrav}) and the metallicity from the [Fe/H] abundance.
Since late-type giant-star atmospheres may be out of LTE
excitation balance for low excited Fe lines \cite{allende,steenbock85},
these lines could not be used to determine
the effective temperature (e.g. Thor\'en 2000)\nocite{thoren00calletter}.
Therefore literature values for \Teff \ were used through the whole
analysis and furthermore the Fe abundance was determined by the use of
highly excited lines only ($\chi > 3.5$ eV.

\section{Model atmospheres and abundance analysis}

\subsection{Model atmospheres}

A version (Asplund et al. 1997)\nocite{asplund_marcs} of the LTE,
plane-parallel, flux constant MARCS atmospheres
(Gustafsson et al. 1975)\nocite{gustafsson_marcs}
were used to represent the atmospheres of the observed stars.

\subsection{Analysis codes}

The Uppsala analysis codes SPECTRUM and EQWIDTH were used for the
abundance derivation. SPECTRUM
was used to fit synthetic spectra to observed ones.
The equivalent widths of the lines fitted were analysed
with EQWIDTH, which yields mean values and scatter of
abundances from measured equivalent widths.

\subsection{Atomic and ionic lines}

Preliminary atomic line data was fetched from the VALD database \cite{vald}.
The intention of our work was to make a differential analysis
relative to the Sun; consequently
we fitted the oscillator strengths of the lines to match
the strength of the solar flux spectra that were taken during the
observational runs.
The VALD oscillator strengths  usually agree within one
order of magnitude with the astrophysical oscillator strengths.
For some lines the solar lines were too weak to enable a
reliable value of the oscillator strength.
In these cases values from the literature were used.
The adopted atomic line data is given in Table\,1
(available in electronic form only) which gives the wavelength, $\log gf$ value,
excitation energy, and damping parameters for each of the 159 lines of atoms
and first ions with wavelengths between 4388 and 8758\,\AA.
The species are identified in Table\,\ref{abund1}, Sect\,\ref{abund3}.

A flux-constant solar model may not be ideal for determination
of oscillator strengths.
The present work is a differential study which means that most errors in
the atomic data will hopefully cancel out when data of solar-like stars are
interpreted.
Most stars in this sample are not dwarf-like, rather they
are subgiants and giants.
The effect of errors in the astrophysical $\log gf$ values would therefore be
expected to show up as small but systematic shifts in the derived abundances.
The shape of these patterns are, however, not expected to change very much
due to these errors.

Pressure broadening data for the lines was based on quantum mechanical
calculations by
Barklem \& O'Mara (1998, and references therein)\nocite{barklem98:3}.
For many lines, particularly ionic lines, such data are not available.
In those cases the classical Uns\"old damping constant was assumed and
multiplied by a factor typical for the element.
These factors are listed in Table\,1 (available in electronic form only).
For the Cu\,{\sc i} line at 5105.5\,\AA, hyperfine structure splitting was
considered and the solar isotopic ratio $^{63}$Cu/$^{65}$Cu\,=\,0.69/0.31
was assumed for all stars.

First, all stars were analysed by means of synthetic spectroscopy
in the regions centered at 4383, 5269 and 6162\,\AA.
The results from these regions gave new values for metallicity (Fe abundance),
surface gravities (pressure broadening of strong lines) and microturbulence
parameters (balance in abundances from strong/weak Ca, Fe and Ni lines).
The remaining regions were analysed with atmospheric models based on these
parameters.

During the analysis up to 200 different atomic, ionic and molecular features
were analysed, and in many cases dropped if close inspection showed that they
are blended by unknown lines (e.g., in the solar spectrum or exposed when a
particular line gives deviating abundances for atmospheres with low \Teff).

\subsection{Molecular lines}

Absorption features from CH, C$_2$, CN, $^{13}$CN and MgH present in our
wavelength regions were used for abundance determinations of carbon, nitrogen
and magnesium.
Molecular lines were detected in most of our programme stars except for the
hottest ones.
The ``molecular'' abundances are presented separately from those determined
from atomic and ionic lines
to enable discussion of the internal consistency in the analysis.
Line data for the most important molecular features; wavelengths, excitation
energies and oscillator strengths, are given in Table\,2
(only available in electronic form).

Adopting the oxygen abundances from Sect.\,\ref{oxygen},
carbon abundances were found from CH and C$_2$ lines, and
nitrogen abundances were derived from CN lines.
For stars without spectra of oxygen lines, we adopted oxygen abundances typical
for stars of EAGLNT with similar [Fe/H].

Basic data for $^{12}$CH A--X (0,0) and (1,1) lines in the wavelength regions
$\lambda\lambda$ 4180--4220 and 4285--4315 were adopted from
J{\o}rgensen et al. (1996)\nocite{jorgensen}.
Features of the A--X (1,1) band were shifted to the solar wavelengths reported by
Moore et al. (1966)\nocite{mooretab}, and all $gf$ values were decreased by
0.12\,dex.
This shift would not have been necessary if the semiempirical solar model of
Holweger-M\"uller (1974)\nocite{holweger} had been adopted.
For consistency reasons we here use a MARCS theoretical model also for
the Sun, hoping that possible model deficiences will cancel out to some degree
in a relative abundance analysis.
A maximum of 22 useful CH features were used for each star and the line-to-line
abundance scatter was typically 0.05\,dex (ranging between 0.03 to 0.10).

\setcounter{table}{2}
\begin{table*}\caption{\label{grav}
Surface gravities from different lines and methods.
The parallax $0\fs 045$ was adopted for HD\,82734
(van Altena \& Hoffleit 1996).
Fe (ion bal.) shows gravities assuming LTE ionisation balance for Fe.
The column marked ``Adopted'' shows the surface gravities used for the
abundance determination}
\begin{tabular}{l c c c c c c c c c }
Star & Mg\,{\sc i}& Mg\,{\sc i} & Ca\,{\sc i} & Fe\,{\sc i} & Fe\,{\sc i} &
                                               Fe & {\sc Hipparcos} & Adopted\\
           & 5167 & 5172 & 6162 & 4383 & 5269 & ion bal. & from $\pi$ \\
\hline
HD\,400    & 4.13 & 3.98 &      & 4.15 &      & 3.85 & 4.07 & 3.98  \\
HD\,2151   & 3.78 & 3.70 & 4.08 &      & 3.83 & 3.87 & 3.95 & 3.93  \\
HD\,6734   & 3.70 & 3.70 & 3.40 &      & 3.55 & 3.28 & 3.47 & 3.40  \\
HD\,18907  & 4.31 & 4.31 & 4.01 &      & 4.01 & 3.90 & 3.86 & 4.21  \\
HD\,22484  & 4.00 & 4.00 & 3.96 &      & 4.12 & 3.92 & 4.09 & 3.96  \\
HD\,23249  & 3.65 & 3.60 &      &      &      &      & 3.67 & 3.82  \\
HD\,24616  & 3.50 & 3.30 & 3.10 &      & 3.10 & 3.33 & 3.22 & 3.10  \\
HD\,35410  & 3.00 & 2.86 & 3.16 &      & 3.10 & 2.85 & 3.01 & 3.16  \\
HD\,40409  &      & 3.40 & 3.30 &      & 3.30 & 2.62 & 3.12 & 3.30  \\
HD\,61421  & 4.30 & 4.15 &      &      &      & 3.80 & 4.02 & 4.02  \\
HD\,62644  & 3.50 & 3.42 & 3.50 &      & 3.75 & 3.27 & 3.65 & 3.42  \\
HD\,82734  &      &      & 3.25 &      &      & 2.33 &      & 3.11  \\
HD\,111028 &      &      & 3.10 &      & 3.10 & 2.55 & 3.03 & 3.10  \\
HD\,115577 & 2.90 & 2.75 & 2.75 &      & 2.65 & 2.45 & 2.30 & 2.75  \\
HD\,130952 & 2.45 & 2.28 & 2.18 &      & 2.43 & 1.50 & 2.43 & 2.43  \\
HD\,142198 & 2.76 & 2.61 & 3.11 &      &      & 2.25 & 2.57 & 2.91  \\
HD\,168723 & 3.30 & 3.10 & 2.95 &      & 2.95 & 2.87 & 3.01 & 2.95  \\
HD\,196171 & 2.70 & 2.60 & 3.00 &      &      & 2.10 & 2.63 & 3.00  \\
HD\,197964 & 3.10 & 2.95 & 3.25 &      & 3.25 & 2.75 & 3.10 & 3.25  \\
HD\,199623 & 3.97 & 3.97 & 4.27 &      &      & 3.95 & 4.13 & 4.27  \\
HD\,207978 & 4.05 & 3.90 &      & 4.05 &      & 3.90 & 4.04 & 3.90  \\
HD\,212487 & 3.75 & 3.75 &      & 3.60 &      & 3.60 & 3.83 & 3.60  \\
HD\,219617 & 4.10 & 4.05 &      & 3.90 &      & 3.90 & 3.83 & 3.90  \\
\hline
\end{tabular}
\end{table*}

$^{12}$C$^{12}$C A--X (0,0) line data in the 5085--5165\,\AA\ region was adopted
from Querci et al. (1971)\nocite{querci}.
The wavelengths of the dominating lines were modified to fit
our solar observations and the solar flux atlas of
Kurucz et al. (1984)\nocite{solaratlas}.
No modification of the $gf$ values was found to be necessary.
Carbon abundances for 17 stars were derived from typically 3, but depending
on the observed wavelength regions up to 7, C$_2$ features that were judged
to be negligibly contaminated by other lines.
The line-to line abundance scatter was typically 0.03\,dex (0.00 to 0.09).

Violet and red $^{12}$C$^{14}$N line data in the 4180--4215 (B--X (0,1) and
(1,2)) and 7980--8020\,\AA\ (A--X (2,0))
regions were adopted from Plez (2001), with the experimental CN
dissociation energy of 7.77\,eV of Costes et al. (1990)\nocite{costes}.
The CN line wavelengths were found to give a very good agreement with the
solar spectrum.
In the 4200\,\AA\ region the oscillator strengths had to be decreased by
0.12\,dex, just like for the CH lines in the same region discussed above.
In the red region the solar spectrum was very well fit without any
modifications to the CN line data.
Nitrogen abundances were derived from up to 25 lines in the violet region,
with line-to-line scatter of typically 0.06\,dex.
In the red region up to 8 lines could be used with a similar line-to-line
scatter.

In the 7920--8020\,\AA\ region 10--20 red $^{13}$C$^{14}$N A--X (2,0)
features could be identified and used for determinations of $^{12}$C/$^{13}$C
isotope ratios from the observations of 6 stars.
This measurement is insensitive to uncertainties in the abundances of C, N
(and O) since only the relative strengths of the $^{12}$CN and $^{13}$CN
features enter.
Line data from Plez (2001) were adopted, and for the about 100 strongest lines
laboratory wavelengths were found in Wyller (1966)\nocite{wyller} and adopted.
Wavelength corrections of +0.1 to +0.4\,\AA\ were typically needed to Plez's
data.
The strongest $^{13}$CN feature in the region falls at 8004.3--8004.7\,\AA\
and this was the main criterion for the isotopic ratio, with consideration
also of the weaker features.
The uncertainties were conservatively assumed as the standard deviation
(scatter) of the various $^{13}$CN features used in each spectrum.
The resulting $^{12}$C/$^{13}$C ratios are presented in Table\,\ref{abund1}.
Spectra for another six stars were obtained for the 7970\,\AA\ region.
Four of these were too hot to show any CN lines (HDs 400, 61421, 207978
and 212487), and those for the other two (HDs 6734 and 24616) ended
bluewards of the 8004\,\AA\ $^{13}$CN feature.

MgH A--X (0,0) line data for the region 5140--5195\,\AA\ was adopted from
Kurucz (1994).
Wavelengths and $gf$ values were then individually modified to fit
our solar observations and the solar flux atlas of Kurucz et al. (1984).
For 9 stars our spectra enabled abundance determinations from typically
7 MgH lines, with line-to-line scatters of typically 0.10\,dex.
The derived magnesium abundances are not sensitive to the C,N,O
element abundances within their uncertainties.

\section{Surface gravities}
\label{surfgrav}

The purpose of this investigation is to study how useful
subgiants may be for chemical evolution studies.
The luminosity class alone is usually not sufficient to identify such objects,
$\log g$ is also needed.
Spectroscopic determination of surface gravity is a powerful
distance-independent method to sort out subgiants from giants or dwarfs.
Another method is by the use of accurate parallaxes.

\subsection{Spectroscopy}

Pressure-broadened lines are sensitive to surface gravity and we use the
wings of such lines to derive $\log g$ (Blackwell \& Willis 1977,
Edvardsson 1988) \nocite{blackwellwillis,edvardsson88}. Ideally, to
compensate for uncertainties in the model atmospheric parameters and
possible deviations from LTE, the abundances used in the calculation of
the strong line should be derived from weak lines with the same lower
energy level.

Depending on the stellar metallicity, we observed some of the spectral regions
containing the following strong and pressure-broadened metal lines:
Fe\,{\sc i} 4383.6 and 5269.5\,\AA, Mg\,{\sc i} 5167.3 and 5173.7\,\AA\
and Ca\,{\sc i} 6162.2\,\AA.
The hydrogen pressure broadening parameters of these lines were adopted
from Anstee \& O'Mara (1995).
Derived $\log g$ values are presented in Table\,\ref{grav}.
The surface gravities finally adopted in the analysis are
not necessarily the straight mean of the strong-line gravities, since
higher weight was put on lines with superior line-wing fits.
For instance, some line wings were so wide that the continuum placement became
difficult and some were blended with other strong lines.
The strong Fe\,{\sc i} lines were used with the abundance derived
only from lines with low excitation energies (which were not used for the
abundance analysis in Sect.\,\ref{abund3}), while for the strong Mg and Ca
lines we adopted the general abundances given later in Table\,\ref{abund1}.
Either, never both, of the two strong Fe\,{\sc i} lines were used for a star.
Gravities derived from assuming Fe ionisation balance
are also presented in the table, but these were not used in this analysis.

\begin{figure}[h]
 \resizebox{0.89\hsize}{!}{\includegraphics{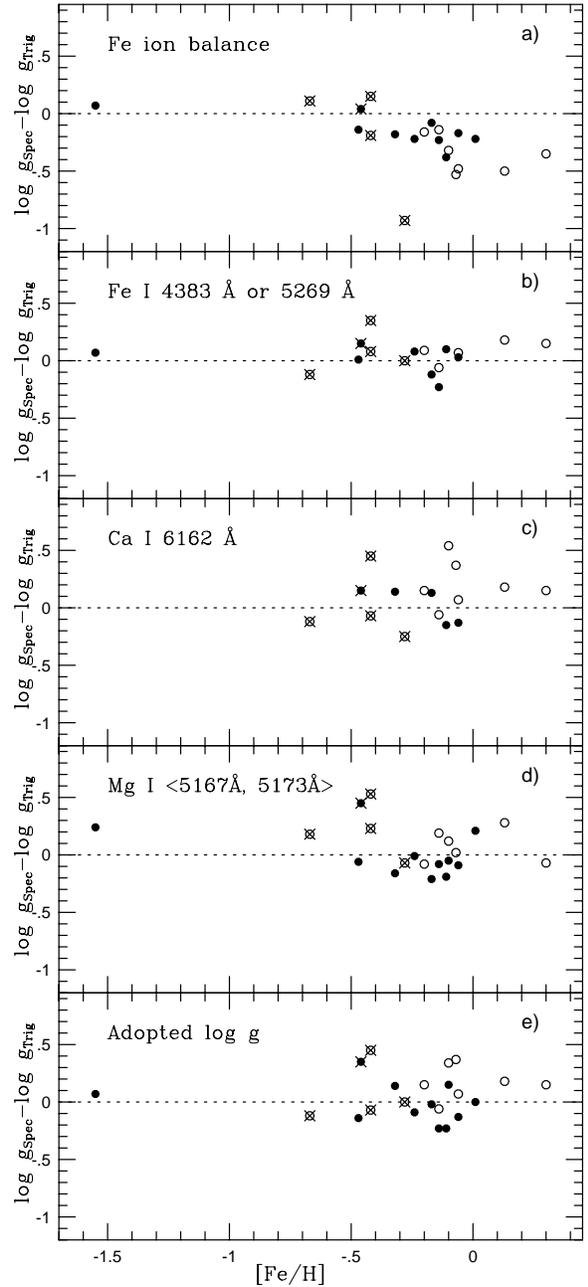}}
  \caption[]{Surface gravities derived by different methods minus parallax
   gravities.
   The "Fe ionization balance" panel, {\bf a}), represents results based on
   Fe\,{\sc i} and Fe\,{\sc ii} lines assuming LTE ionisation balance
   (these were not further used in the analysis).
   For the strong Fe lines, panel {\bf b}), iron abundances derived from
   exclusively low-excitation lines were used.
   The Ca and Mg panels, {\bf c}) and {\bf d}), show results from using
   strong lines with abundances from Table\,\ref{abund1}.
   Panel {\bf e}), "Adopted $\log g$", shows the surface gravities that were
   adopted in the analysis.
   Subgiant-branch stars are shown as solid dots and giants as open circles.
   Candidate thick-disk stars are identified by superposed crosses}
 \label{loggsloggpi}
\end{figure}

\subsection{Parallaxes}
\label{parallaxes}

Trigonometric surface gravities were derived by Eq.\,\ref{eq1}.
\begin{eqnarray}
\label{eq1}
\log \frac{g}{g_{\odot}}=4 \log \frac{T_{\rm eff}}{T_{{\rm eff} \odot}}+
\log \frac{\cal{M}}{\cal{M}_{\odot}} + 0.4(M_{\rm bol}- M_{\rm bol \odot})
\end{eqnarray}
Masses were derived from the theoretical ($T_{\rm eff},L$) isochrones of
Girardi et al. (2000).
One subgiant, HD\,82734, appeared to be extremely luminous, why we consulted
the Yale Parallax Catalogue (van Altena \& Hoffleit 1996) \nocite{yalepar}
and found a more reasonable parallax 5 times larger than that reported
by {\sc Hipparcos} (which is 9.76 mas).
Bolometric corrections were calculated from
Tables 3--6 in Mihalas \& Binney (1981)\nocite{mihalasbinney}.
The resulting trigonometric surface gravities are presented in
Table\,\ref{grav}.

\subsection{Comparison between the methods}

Allende Prieto et al. (1999)\nocite{allende}
noted that large differences between $\log g$ values derived from
parallaxes and LTE Fe ionisation balance do exist and can not
yet be explained.
The agreement between our gravities from the LTE ionisation balance
method and those from parallaxes is rather poor, see Fig.\,\ref{loggsloggpi}d).
The ionization balance gravities are systematically lower than the
trigonometric ones for the more metal-rich stars.
This trace of overionisation in Fe suggests that the derivation of $\log g$
values from the ionisation balance may be hazardous.
The pattern displayed by our stars does however not match the predicted NLTE
pattern or the compiled observational data of Fig.\,10 in
Allende Prieto et al. (1999).

In Fig.\,\ref{loggsloggpi} also our ``strong-line'' surface gravities are
compared with the parallax gravities.
The $\log g$ values from strong Fe\,{\sc i} lines agree better
with parallax gravities than any other criterion used.
For Mg there are no low excited lines in the observed spectral regions and the
Ca line wings are not wide enough to be used for metal poor stars.
The adopted surface gravities are weighted mean values of the strong-line
gravities, with weights subjectively assigned depending on the fits to the line
wings.

An alterative way of determining the stellar gravity would be to use the MgH
lines, which are known to be gravity sensitive
(Bell, Edvardsson \& Gustafsson 1985\nocite{BEG}, and references therein),
with the Mg abundance derived from atomic lines.
We have explored whether there is a correlation
between the differences between the Mg abundances of Table\,\ref{abund1} as
derived from MgH and Mg\,{\sc i} lines, respectively, and the differences
between the adopted (spectroscopic) and the trigonometric gravities, and find no
such correlation.
That is, surface gravities from MgH would be consistent with the adopted ones
at a mean, but are not good enough to improve them significantly.

\section{Ages and masses}
\label{agesandmasses}

\begin{figure}[ht]
 \resizebox{\hsize}{!}{\includegraphics{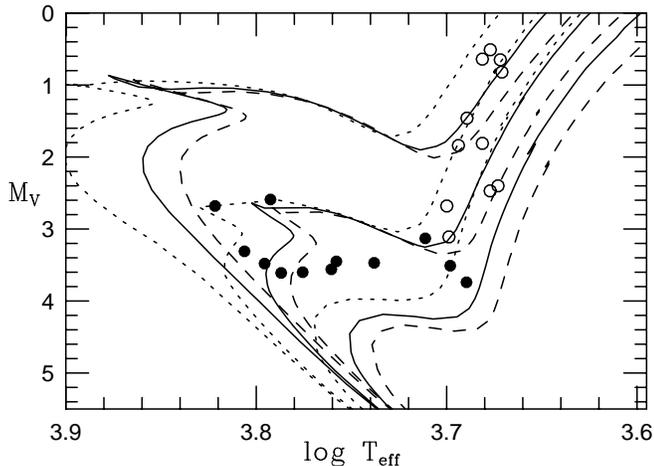}}
  \caption[]{The positions of our programme stars among isochrones of
   Girardi et al. (2000) for three metallicities and three ages. The
   metallicities, [Fe/H], are $+0.20$ (dotted lines), $0.00$ (solid lines)
   and $-0.38$ (dashed lines), and the ages are, from upper left to lower
   right; 1.12, 3.16 and 12.6 Gyr. The subgiant stars are shown as filled
   circles and the giants by open circles}
 \label{isochrones}
\end{figure}

To estimate ages and stellar masses we used isochrones from
Girardi et al. (2000). These authors adopted magnitudes in the $V$ band
from Kurucz (1992) model atmospheres \nocite{kurucz1992}.
Interpolation in three dimensions were done with MATLAB to
find age and mass for stars with metallicities between the isochrones
available in the set.
Fig.\,\ref{isochrones} shows our programme stars with selected isochrones
from Girardi et al.

We compared the Girardi et. al. ages with ages derived from
the isochrones by Bergbusch \& VandenBerg (2001) \nocite{bergbusch01} and
those by
Yi et al. (2001) \nocite{yy} and found no significant differences.

The resulting masses and ages are shown in Table\,\ref{abund1}.
For stars overlapping with Edvardsson et al. (1993)
the ages derived agree very well. The masses and ages of the
latter sample of stars were rederived by Ng \& Bertelli (1998)\nocite{ng}
and these also agree with our derived corresponding results.

The possibility that parallax errors will make old horisontal branch or clump
stars to be mistaken for young giant stars was investigated.
This is most likely to occur for approximately solar metallicity.
Three stars, HD\,35410, HD\,115577 and HD\,130952
are within one $\sigma$ in $M_V$ (using the {\sc Hipparcos} standard error as
$\sigma_\pi$) and one $\sigma$ in \Teff (150 K) from the {\it old} clump in
the isochrones corresponding to that metallicity  ($Z \sim 0.019$).
These three may therefore possibly be core-helium-burning clump stars.

\section{Dynamical aspects}
\label{dynamics}

\begin{figure}[ht]
 \resizebox{\hsize}{!}{\includegraphics{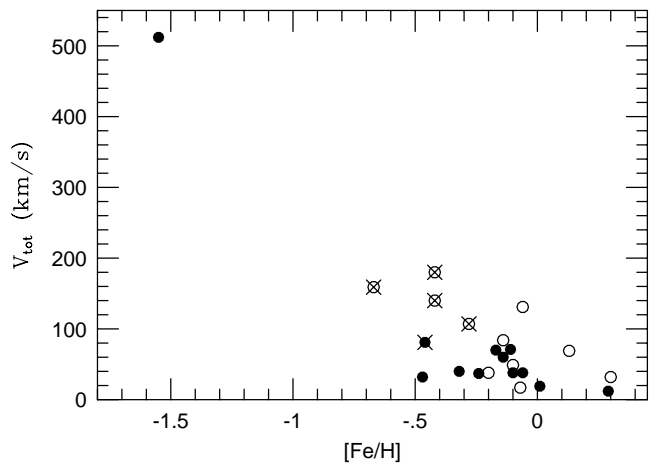}}
  \caption[]{[Fe/H] vs. total velocity relative to the LSR.
   Subgiant-branch stars are shown as solid dots and giants as open circles.
   Candidate thick-disk stars are identified by superposed crosses}
 \label{vtotfe}
\end{figure}

All objects have been observed by the {\sc Hipparcos} satellite
and have accurate proper motion and parallax data.
The exception is HD\,82734 as discussed in Sect.\,\ref{parallaxes}.
The radial velocities were taken from the SIMBAD database, except for the
radial-velocity variable HD\,18907 for which a mean systemic value of
$RV=+42.7$\,km/s was adopted \cite{nidever}.
Galactic $U$ (positive in the direction of the galactic center), $V$
(positive in the direction of the rotation of the galactic disk) and
$W$ (positive towards the north galactic pole) velocities were
calculated by the formalism given by Johnson \& Soderblom (1987).
The total space velocity relative the local standard of rest
(Dehnen \& Binney 1998) is plotted versus iron abundance in Fig.\,\ref{vtotfe}.

In this figure and in most of the following ones we have identified five
candidate thick-disk stars by overcrossed symbols.
These were identified by a combination of kinematical and chemical signatures.
Following Bensby et al. (2003)\nocite{bensby} local stars
with total space velocities relative to the local standard of rest larger than
about 75\,km/s have large probabilities of belonging to the ``thick'' component
of the galactic disk.
Our sample contains eight stars fulfilling this criterion.
HD\,219617 has a retrograde galactic orbit and a our lowest metallicity
([Fe/H]\,$=-1.55$) which puts it in the halo-star population.
Four giants, HD\,6734, HD\,24616, HD\,115577 and HD\,130952, and one subgiant,
HD\,18907, have peculiar velocities between 80 and 180\,km/s,
and overabundances of aluminium and $\alpha$-elements as compared to the majority
of disk stars of similar metallicities which is our ``chemical''
criterion for belonging to the thick-disk
(following e.g., Fuhrmann 1998\nocite{fuhrmann} or
Feltzing et al. 2003\nocite{feltzingetal}).
These five are represented by crossed symbols in the figures throughout this article.
Two more giants, HD\,111028 and HD\,168723 have peculiar velocities of 131 and
84\,km/s, respectively.
HD\,111028 with [Fe/H]\,$=-0.06$ also shows high abundances of the $\alpha$
elements S, Si and Ca, but not of O and Ti.
HD\,168723 with [Fe/H]\,$=-0.14$ shows similar but even milder chemical
peculiarities.
These two giants may be metal-rich thick-disk star of the kind discovered by
Feltzing et al. (2003)\nocite{feltzingetal}, but we have
chosen not to emphasize them further here.

The five thick-disk candidates are discussed further  in Sects.\,\ref{alpha}
and \ref{thickdisk}.

\section{Errors}
\label{errors}

\subsection{Abundances}
\label{abund2}

The chemical abundances are sensitive to
the accuracy of stellar parameters, to the synthetic spectral fitting
to observational data, to the atomic and molecular data
and to errors caused by the assumptions of LTE and mixing-length convection
in plane-parallel geometry.

\subsubsection{Stellar parameters}
\label{parerrors}

Accurate stellar parameters are essential for getting reliable abundances.
The effects of such errors on the abundances are reviewed in this section.
Uncertainties in these parameters affect not only the abundances but also each
other, e.g. $\log g$ depends on the adopted \Teff.

\begin{figure}[ht]
 \resizebox{\hsize}{!}{\includegraphics{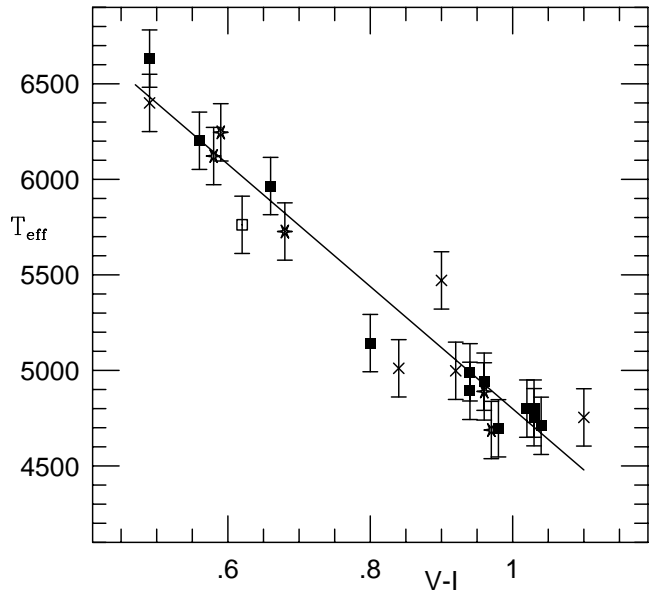}}
  \caption[]{\Teff versus $V-I$ colours. Open square: [Fe/H] $= -1.55$,
   crosses: $-0.7 <$[Fe/H] $< -0.4$,  asterisks:  $-0.4 <$[Fe/H] $< -0.15$,
   filled squares: [Fe/H] $> -0.15$. The line is fitted to the 22 stars with
   [Fe/H] $> -1.0$. These 22 scatter by 163\,K around the line.
   The error bars represent our estimated \Teff uncertainty of 150\,K}
 \label{teffvi}
\end{figure}

Effective temperatures are often derived spectroscopically by requiring that
the abundances of iron derived from Fe\,{\sc i} lines of different excitation
should not differ systematically. We note a systematic deviation for
Fe abundances from low excited lines in our low $\log g$
(and low $T_{\rm eff}$) stars.
This has earlier been suggested to be an NLTE effect for giant stars
\cite{steenbock85} and following their results we thus choose to use high
excited lines for our Fe abundances.
The effective temperatures were compiled from earlier analyses, and they are
therefore not very homogeneous.
We estimate from earlier experience and the scatter between different
investigations that our \Teff determinations are internally consistent within
about 150\,K.
As a simple test we plot in Fig.\,\ref{teffvi} our adopted values of \Teff vs.
$V-I$, as given in the {\sc Hipparcos} catalogue (ESA 1997).
Neither empirical nor theoretical model-atmosphere \Teff calibrations
suggest any appreciable metallicity sensitivity of the $V-I$ colour at
effective temperatures above 4000\,K.
Nevertheless, since line blanketing may affect the relation between \Teff and
$V-I$, we have separated the stars into four different metallicity intervals,
and plot the stars in each interval with different symbols and an error bar of
$\pm 150$\,K.
22 of our 23 stars have $-0.7\le$\,[Fe/H]\,$+0.3$.
The remaining star, HD\,219617 (marked by an open square in Fig.\,\ref{teffvi})
has [Fe/H]\,$=-1.55$ and falls below the line in the diagram.
The scatter seen relative to a mean relation for the stars
is consistent with an uncertainty in \Teff of about 150\,K.
These considerations make us adopt an uncertainty of $\pm$\,150\,K for the
stellar effective temperatures for the disk-metallicity stars.

\begin{figure}[ht]
 \resizebox{\hsize}{!}{\includegraphics{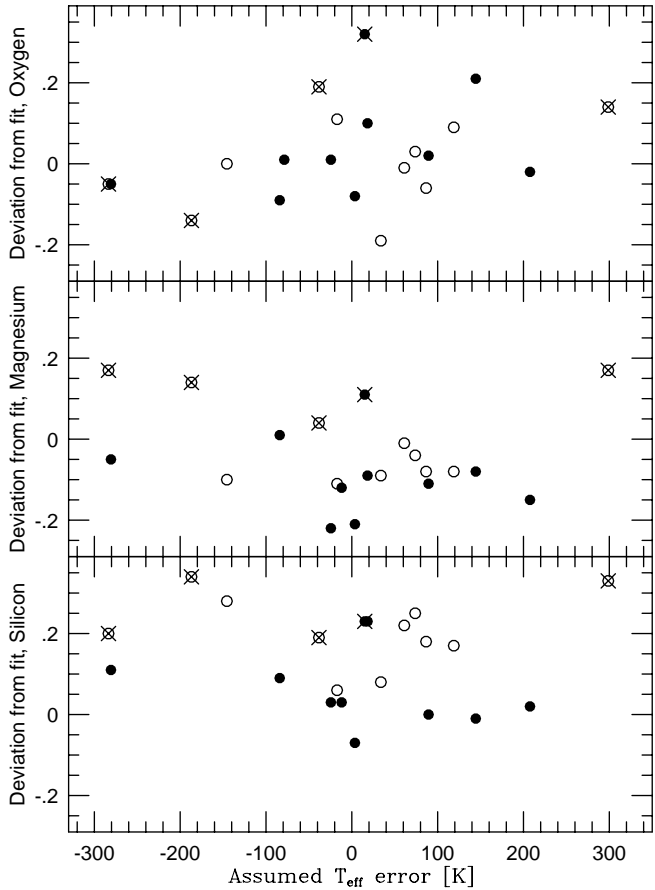}}
  \caption{Abundance deviations from fit to Edvardsson et al. (1993) data
   versus \Teff deviations from the line in Fig.\,\ref{teffvi}, for stars with
   [Fe/H]\,$> -1.0$. Symbols as in Fig.\,\ref{vtotfe}}
 \label{bdpteffvi}
\end{figure}

The possibility that errors in the effective temperatures is an important
source of the scatter in our abundance-abundance diagrams
(cf. Fig.\,\ref{allab}, Sect.\,\ref{abund3}) was empirically tested:
If errors in our \Teff determinations were dominating the
scatter seen in our abundance diagrams we would
expect these abundance errors to be correlated with the \Teff errors.
To test this we assumed the \Teff deviations from the line in
Fig.\,\ref{teffvi} to be our \Teff errors.
We next adopted the difference between our [X/Fe] and a linear fit of
[X/Fe] vs. [Fe/H] from EAGLNT (for our 22 stars with [Fe/H]\,$>-1.0$).
These differences for O\,{\sc i}, Mg\,{\sc i} and Si\,{\sc i} were plotted vs.
the assumed \Teff errors, but no appreciable correlations could be found,
see Fig.\,\ref{bdpteffvi}.
Errors in our effective temperatures are thus unlikely to dominate the scatter
in the abundance diagrams.

The subgiant HD\,18907 has recently been presented to show radial velocity
variations \cite{setiawan}.
First estimates of the orbital parameters gives $P \approx 9700$\,days,
$a \approx 10$\,AU and a mass function $f(m)=0.018$.
\cite{udryprivate,setiawanprivate}.
Our derived primary mass is 1.1\,$M_\odot$ and a $90\degr$ orbital
inclination gives a minimum secondary mass of 0.32\,$M_\odot$.
Assuming an inclination of $60\degr$ gives a mass of $m_2=0.4\,M_\odot$
for the secondary, corresponding to a M2 dwarf with $M_V \approx +9.9$
\cite{cox}.
A parallax of 33\,mas \cite{hipparcos} then gives a secondary flux of 0.3\%
of that of the primary in the $V$ passband and a separation on the sky of a
fraction of an arcsecond.
Nevertheless, without information of the orbital inclination there is still a
50\% chance of a more massive secondary and there remains a possibility of
errors in the equivalent widths for HD\,18907 due to multiplicity.
At the time of observation the primary had a radial velocity +1.3\,km/s higher
than the systemic velocity and the secondary should therefore be blue-shifted
by more than this amount relative to the primary
($-6$\,km/s for $m_2=0.4\,M_\odot$).
We can not detect any traces of the secondary in our spectra for HD\,18907.

Setiawan et al. (2004)\nocite{setiawan2004} also report that HD\,62644 is a
binary with $m_2 \sin i=0.32\,M_\odot$.
Also for this star we can not find any trace of a secondary spectrum, and
the abundance results for the star give no reason to suspect any influence of
the secondary stellar spectrum.

In Table\,\ref{graverr} the sensitivities of $\log g$ due to uncertainties
in \Teff for three typical programme stars are presented.
For spectroscopic gravities, \Teff affects the fraction of neutral atoms,
as well as the continuous opacity, giving generally weaker lines for
higher $T_{\rm eff}$.
The trigonometric surface gravity is affected
through the \Teff contribution to Eq.\,\ref{eq1}
(also including a change in bolometric correction).
Note that the sensitivity to \Teff errors is rather
small for the parallax method.
The strong-line method for determining gravities is also quite insensitive to
errors in \Teff (Blackwell \& Willis 1977, Edvardsson 1988) when the metal
abundance is derived from weak lines with similar excitation energies.
This is the case for the strong Ca and Fe lines in our sample, while for the Mg
triplet lines (2.7\,eV) we had to use weak lines excited to between 5 and 6\,eV
for the abundance determinations.

\begin{table*}\caption[]{\label{graverr}
The sensitivities of surface gravities, $\log g$, derived from line-wings and
parallaxes to a temperature change $\Delta T_{\rm{eff}}=+150$\,K}
\begin{tabular}{l c c c c c c }
Star       & Fe\,{\sc i} & Mg\,{\sc i} & Mg\,{\sc i} & Fe\,{\sc i} & Ca\,{\sc i} & {\sc Hipparcos} \\
           & 4383\,\AA\  & 5167\,\AA\  & 5172\,\AA\  & 5269\,\AA\  & 6162\,\AA\  & $\pi$     \\
\hline
HD\,400    & +0.00       & +0.15       & +0.15       &             &             & +0.05     \\
HD\,18907  &             & +0.20       & +0.20       & +0.00       & +0.00       & +0.05     \\
HD\,130952 &             & +0.35       & +0.35       & +0.10       & +0.00       & +0.08     \\
\hline
\end{tabular}
\end{table*}

The precision of the parallax method is only weakly depending on errors in
temperature and masses.
The resulting mean difference
between the Ca\,{\sc i} 6162\,\AA\ gravities and the parallax gravities is
$+0.10$\,dex with a standard deviation of $0.22$\,dex.
For the Fe\,{\sc i} lines the corresponding numbers are $+0.05$ and $0.13$\,dex.
The mean difference between the adopted gravities in the analysis and those of
the parallax method is $+0.06$ with a scatter of $0.19$\,dex.
We conservatively judge the mean error in the adopted surface gravities to be
about 0.2\,dex.

The effects on the derived abundances due to errors in \Teff and $\log g$ are
shown in Table\,\ref{abparerr} for three stars with typical parameters.
Our estimated error of 0.2\,dex in $\log g$ would produce a typical abundance
effect of $\sim 0.05$\,dex and a \Teff error of 150\,K changes the abundances
$\sim 0.10$\,dex

The microturbulence parameters were derived by requiring strong and weak Fe
lines to yield equal abundances.
The uncertainty in abundances due to this error source was tested for the stars
in Tables\,\ref{graverr} and \ref{abparerr}.
It was found to be significant only for the metal-rich cool giant HD\,130952.
For most [X/Fe] abundance ratios a change in $\xi_{\rm t}$ of 0.5\,km/s caused
an [X/Fe] shift of typically $\sim 0.08$\,dex for this star.
For the other stars an equal uncertainty in $\xi_{\rm t}$ is less important
for the abundances, typically $< 0.05$\,dex.

\begin{table*}\caption{\label{abparerr}
Typical sensitivities of derived abundances, [X/H], to changes in model \Teff
and $\log g$}
\begin{tabular}{l r r c r r c r r}
\noalign{\smallskip}
Star                 &\multicolumn{2}{c}{HD\,400 \hfill SGB}  &
                                 &\multicolumn{2}{c}{HD\,18907 \hfill SGB} &
                                                         &\multicolumn{2}{c}{HD\,130952 \hfill RGB} \\
$T_{\rm eff}$/$\log g$/[Fe/H]
                     &\multicolumn{2}{c}{6122/3.98/$-$0.24} &
                                 &\multicolumn{2}{c}{5271/4.21/$-$0.46} &
                                                         &\multicolumn{2}{c}{4688/2.43/$-$0.28} \\
\noalign{\smallskip}
                     \cline{2-3}           \cline{5-6}           \cline{8-9}
\noalign{\smallskip}
                     &$\Delta T_{\rm eff}$&$\Delta \log g$ &
                                 &$\Delta T_{\rm eff}$&$\Delta \log g$ &
                                                         &$\Delta T_{\rm eff}$&$\Delta \log g$\\
                     &+150\,K&$-$0.3\,dex& &+150\,K&$-$0.3\,dex& &+150\,K&$-$0.3\,dex\\
\hline
Li (Li\,{\sc i})     &    0.11 &    0.01 & &    0.15 &    0.01 & &    0.21 &    0.00 \\
C ($[$C\,{\sc i}$]$) &         &         & & $-$0.07 & $-$0.16 & & $-$0.15 & $-$0.23 \\
C (CH)               &         &         & &    0.11 & $-$0.05 & & $-$0.04 & $-$0.11 \\
C (C$_2$)            &         &         & &    0.06 & $-$0.03 & & $-$0.03 & $-$0.10 \\
N (CN)               &         &         & &    0.10 &    0.01 & &    0.00 & $-$0.03 \\
O ($[$O\,{\sc i}$]$) &         &         & &    0.02 & $-$0.14 & & $-$0.06 & $-$0.19 \\
Na (Na\,{\sc i})     &    0.06 &    0.00 & &    0.09 & $-$0.02 & &    0.11 &    0.00 \\
Mg (Mg\,{\sc i})     &    0.05 &    0.02 & &    0.05 &    0.03 & &    0.03 &    0.01 \\
Mg (MgH)             &         &         & &         &         & & $-$0.04 &    0.16 \\
Al (Al\,{\sc i})     &    0.05 &    0.00 & &    0.08 &    0.02 & &    0.10 &    0.00 \\
Si (Si\,{\sc i})     &    0.04 &    0.00 & &    0.01 & $-$0.01 & & $-$0.05 & $-$0.06 \\
Si (Si\,{\sc ii})    & $-$0.10 & $-$0.10 & & $-$0.14 & $-$0.13 & & $-$0.26 & $-$0.16 \\
S (S\,{\sc i})       &         &         & & $-$0.11 & $-$0.09 & & $-$0.21 & $-$0.12 \\
K (K\,{\sc i})       &    0.13 &    0.08 & &    0.14 &    0.12 & &    0.19 &    0.08 \\
Ca (Ca\,{\sc i})     &    0.09 &    0.03 & &    0.12 &    0.09 & &    0.15 &    0.04 \\
Sc (Sc\,{\sc ii})    &    0.03 & $-$0.11 & &    0.00 & $-$0.12 & & $-$0.02 & $-$0.13 \\
Ti (Ti\,{\sc i})     &    0.13 &    0.01 & &    0.16 &    0.01 & &    0.25 &    0.00 \\
Ti (Ti\,{\sc ii})    &    0.03 & $-$0.11 & &         &         & &         &         \\
V (V\,{\sc i})       &         &         & &    0.19 &    0.00 & &         &         \\
Cr (Cr\,{\sc i})     &    0.07 &    0.00 & &    0.09 &    0.01 & &    0.12 &    0.01 \\
Cr (Cr\,{\sc ii})    & $-$0.01 & $-$0.11 & & $-$0.04 & $-$0.12 & &         &         \\
Fe (Fe\,{\sc i})     &    0.08 &    0.01 & &    0.08 &    0.00 & &    0.07 & $-$0.03 \\
Fe (Fe\,{\sc ii})    & $-$0.01 & $-$0.11 & & $-$0.06 & $-$0.13 & & $-$0.15 & $-$0.16 \\
Co (Co\,{\sc i})     &         &         & &    0.10 & $-$0.02 & &    0.09 & $-$0.04 \\
Ni (Ni\,{\sc i})     &    0.10 &    0.00 & &    0.08 & $-$0.02 & &    0.06 & $-$0.05 \\
Zn (Zn\,{\sc i})     &         &         & & $-$0.04 & $-$0.08 & & $-$0.12 & $-$0.11 \\
Y (Y\,{\sc ii})      &    0.06 & $-$0.11 & &         &         & &         &         \\
Nd (Nd\,{\sc ii})    &         &         & &    0.04 & $-$0.12 & &    0.01 & $-$0.13 \\
$\alpha$             &    0.08 &    0.02 & &    0.09 &    0.03 & &    0.10 &    0.00 \\
\hline
\end{tabular}
\end{table*}

\subsubsection{Atomic data}

The scatter (s.d.) in abundances derived from individual spectral lines
around the mean abundance of a species is typically 0.06\,dex.
The formal mean error is naturally smaller, since many of the elements
provided several lines for use.
Tables\,1 and 2 (electronically available) identifies the lines and features
used for each species.

For most atomic lines the oscillator strengths were determined by fitting of
solar spectra.
The precision in $\log gf$ values is then dependent on the solar fits.
We estimate an error $<0.05$ dex from this source.
Additionally, the MARCS solar model may not be ideal for finding the absolute
$\log gf$ values, even if most of the errors will cancel out for
stars with stellar parameters close to the solar values.
For giants this cancellation may only be partial.

The lines used in the abundance analysis are mostly weak and therefore
uncertainties in the pressure damping parameters have been found to have very
small or no effects on the abundance results.

\subsubsection{Molecules}

Since the CO molecule binds a large fraction of the carbon in cool stars,
the uncertainties in the C,N,O molecular equilibria are interconnected.
Based on the atomic [O\,{\sc i}] oxygen abundance, first the carbon and then the
nitrogen abundance were derived. Therefore any error in the oxygen
abundance will affect also the carbon and nitrogen abundances.
We present here some typical abundance effects based on tests with varied
abundances and model parameters.

A mild variation $\Delta \log {\rm O}$ of the assumed oxygen abundance brings a
typical change $\Delta \log {\rm C}$ in the carbon abundances derived from CH
and C$_2$ lines of $\Delta \log {\rm C} \approx +0.5 \Delta \log {\rm O}$ with
an uncertainty in the constant of 0.15 for disk stars cooler than about 5200\,K.
For warmer stars and our halo-metallicity star, the adopted oxygen abundances
have a much smaller influence on the derived carbon abundances.
For our nitrogen abundances derived from lines of CN we find correspondingly
$\Delta \log {\rm N} \approx - \Delta \log {\rm C}$ since the calculated CN lines
vary in proportion to the product of the total carbon and nitrogen abundances.

An effective temperature increase of 150\,K, with the associated change in
the atomic oxygen abundance results in carbon abundance modifications of
between $-0.05$ to $+0.10$\,dex.
The effect for nitrogen ranges between $0.00$ and $+0.10$\,dex.
Similarly, the abundance effects of a change of the surface gravities by
$-0.2$\,dex are $-0.03$ to $-0.08$\,dex for C, and $0.00$ to $-0.03$\,dex
for N.

\begin{figure}[ht]
 \resizebox{\hsize}{!}{\includegraphics{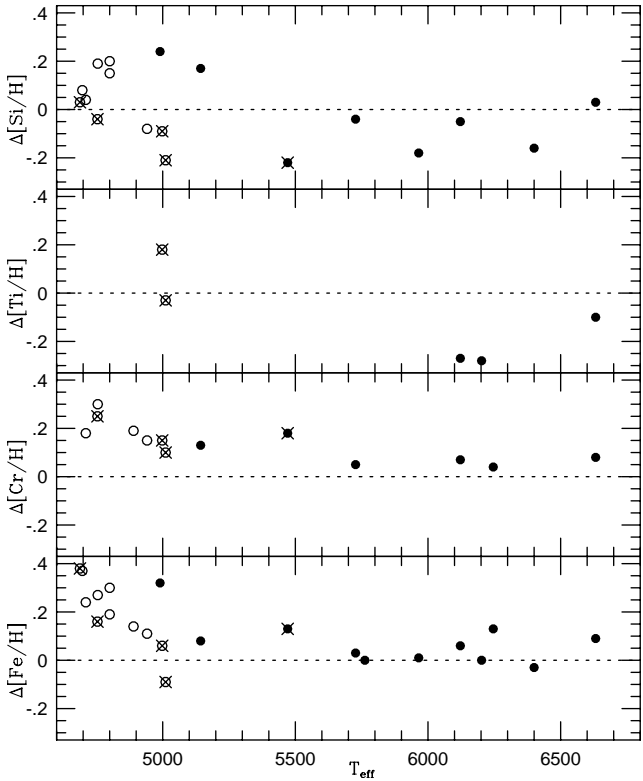}}
  \caption{Differences in abundances derived from lines of ions and neutrals.
   $\Delta$[Fe/H]\,$=$\,[Fe/H]$_{\rm Fe II} -$\,[Fe/H]$_{\rm Fe I}$, etc.
   Solid dots = SGB stars and open circles = RGB stars.
   Five candidate thick-disk stars are identified by crossed symbols}
 \label{overion}
\end{figure}

\subsubsection{Overionization}

In Fig.\,\ref{overion} the empirical ionisation patterns for Si, Ti, Cr and Fe
are plotted against effective temperature.
Our subgiant branch stars are marked as solid dots.
There are obvious trends with effective temperature for the cooler stars
(open circles), which indicate systematic errors in the analysis.
The coolest SGB star, HD\,82734 (\Teff 4990\,K), also shows strong signs of
overionization.
This is also our most metal rich object with [Fe/H]\,$=+0.29$.

\subsubsection{Summary of abundance errors}

From the discussion above we find that the abundances derived are
expected to be correct to about 0.2\,dex for many atomic species,
while the errors in carbon and in particular nitrogen abundances
could be even greater.
These estimates do not account for systematic errors due to departures
from homogeneous (1D) stratification and other uncertainties due to convection.
As has been shown in recent work on the solar CNO abundances
by Allende Prieto, Lambert \& Asplund (2002)\nocite{3D}, such effects may be
considerable (of typically 0.2 dex).
Even if one may hope that the errors in a differential study, like the present
one, are smaller, and that certain abundance ratios are not very sensitive to
these effects, more definitive conclusions must await detailed 3D modelling
of these stars.

\subsection{Uncertainties in ages and masses}
\label{uncertainage}

The age uncertainties depend on the orientation of the isochrones.
For the subgiants these are mostly about 1\,Gyr, while for
the giants they may be several Gyrs, due to uncertainties in the effective
temperatures and the ``vertical'' and densely spaced isochrones.
The effect on the ages of alternatively adopting isochrones from
Bergbusch \& VandenBerg (2001) and Yi et al. (2001)
is typically about 1--2\,Gyr.

A 10\% error in the parallax (or 0.1\,dex in $\log g$) may typically correspond
to about 2 Gyrs in age uncertainty, as derived from the isochrones by
VandenBerg (1983\nocite{Vdb83}, cf. also Gustafsson 1995).
Uncertainties or a cosmic spread of helium abundances of 0.05 in Y
introduce further uncertainties in the ages of typically 1\,Gyr
(as derived from VandenBerg \& Laskarides 1987\nocite{VL}),
while an error in Z of 0.1\,dex may lead to errors of almost as much.
For stars on the sub-giant branch the effects of errors
in convection theory should not be severe on the isochrones
(cf. Pedersen, VandenBerg \& Irwin 1990\nocite{PVI}).
However, the use of stellar models with diffusive
processes (gravitational settling and radiative accelerations)
give reductions of ages of typically 0.5\,Gyr
(Sandage, Lubin \& VandenBerg 2003)\nocite{SLV}.
We conclude that the errors in age estimates may well sum up to 3--4\,Gyrs, at
least for the older stars.
For the RGB stars in the sample, the errors may well be larger, and in
particular the error bars may include very high ages due to the degeneracy of
the isochrones along the giant branch.
For subgiants in the mass interval 0.8--1.5 solar masses, the
age uncertainty corresponds to 0.1--0.2 solar masses in the mass estimate.

\subsection{Uncertainties in the dynamical data}

The {\sc Hipparcos} proper motion data is mostly better than 1 and always
better than 2\,mas/year.
The radial velocity errors are typically 1--2\,km/s, except for
HD\,62644, for which the uncertainty is given as 5\,km/s.
For 15 objects, the uncertainties in the total space velocities are
$\la 2$\,km/s, 6 stars have total errors between 3 and 7\,km/s, while
the giant HD\,115577 is uncertain by 18 and our halo subgiant HD\,219617
with a retrograde galactic orbit has an uncertainty of 102\,km/s in the
total space velocity.
These two stars have small parallaxes and therefore large relative
parallax errors.
In relative measure all but one star have uncertainties
less than 11\% of the total space velocity.
HD\,219617 has error of 20\%.

\section{Abundance results}
\label{abund3}

\begin{table*}
\caption{\label{abund1}
Stellar atmospheric and evolutionary parameters, dynamical properties and chemical abundances.
"Label" indicates whether the star is on the subgiant branch or on the red-giant branch.
The velocities are relative to the LSR and the $U$ velocity component is
directed towards the galactic centre.
The chemical abundances are relative to the Sun:
[X/H]\,$= \log (N_{\rm X}/N_{\rm H})_{\rm Star} - \log (N_{\rm X}/N_{\rm H})_\odot$
and the isotope ratio is given by number of atoms.
Im parenthesis the spectral features used are given.
See Sect.\,\ref{alpha} for definition of the $\alpha$ element abundance.
}
\begin{tabular}{l r r r r r r r r r r}
HD                       &     400  &    2151  &    6734  &   18907  &   22484  &   23249  &   24616  &   35410  &   40409  &   61421  \\
\hline
Label                    &  SGB     &    SGB   & RGB      & SGB      & SGB      & SGB      &  RGB     & RGB      & RGB      &  SGB     \\
\Teff $[$K$]$            &  6122    &  5727    &  4998    &  5471    &  5965    &  4893    &  5011    &  4890    &  4755    &  6632    \\
$\log g$                 &  3.98    &  3.93    &  3.40    &  4.21    &  3.96    &  3.82    &  3.10    &  3.16    &  3.30    &  4.02    \\
$[$M/H$]$                &$-$0.24   &$-$0.17   &$-$0.42   &$-$0.46   &$-$0.06   &$-$0.10   &$-$0.67   &$-$0.20   &   0.13   &   0.01   \\
$\xi_{\rm t}$ $[$km/s$]$ &  1.60    &  2.00    &  1.00    &  1.25    &  2.00    &  2.00    &  1.50    &  1.30    &  1.80    &  1.90    \\
M$_{\rm V}$              &  3.61    &  3.45    &  3.11    &  3.47    &  3.60    &  3.74    &  2.68    &  1.46    &  2.47    &  2.68    \\
Age $[$Gyr$]$            &  6       &  6       &  7       &  6       &  6       & 13       & 11       &  2       &  7       &  2       \\
Mass $[$M$_\odot$$]$     &   1.1    &   1.1    &   1.1    &   1.1    &   1.1    &   1.0    &   1.0    &   1.8    &   1.3    &   1.4    \\
$U$ $[$km/s$]$           &    37    & $-$51    &    60    &    19    &    12    &  $-$4    & $-$15    & $-$25    & $-$62    &    15    \\
$V$ $[$km/s$]$           &  $-$4    & $-$41    &$-$118    & $-$78    & $-$10    &    32    &$-$157    &    24    & $-$28    &  $-$4    \\
$W$ $[$km/s$]$           &  $-$1    & $-$24    &    46    &    12    & $-$35    &    20    & $-$19    &    15    &     9    & $-$11    \\
$V_{\rm tot}$ $[$km/s$]$ &    37    &    70    &   140    &    81    &    38    &    38    &   159    &    38    &    69    &    19    \\
Li (Li\,{\sc i})         &   1.07   &   1.20   &$\le-$0.96&$-$0.39   &   1.24   &          &          &   0.00   &$\le-$1.09&$\le$0.25 \\
C ($[$C\,{\sc i}$]$)     &          &$-$0.14   &          &$-$0.56   &          &          &$-$0.86   &          &$-$0.26   &   0.05   \\
C (CH)                   &$-$0.22   &$-$0.10   &$-$0.40   &$-$0.28   &          &          &$-$0.78   &$-$0.29   &   0.05   &   0.02   \\
C (C$_2$)                &$-$0.24   &$-$0.23   &$-$0.35   &$-$0.39   &$-$0.15   &   0.06   &$-$0.72   &$-$0.30   &   0.05   &          \\
$^{12}$C/$^{13}$C        &          &          &          &          &          &          &          & 20$\pm$5 & 35$\pm$10&          \\
N (N\,{\sc i})           &          &   0.06   &          &          &          &          &          &          &          &   0.10   \\
N (CN)                   &$-$0.41   &$-$0.20   &$-$0.35   &$-$0.58   &          &          &$-$0.86   &$-$0.04   &   0.29   &          \\
O ($[$O\,{\sc i}$]$)     &          &$-$0.24   &$-$0.12   &$-$0.01   &$-$0.06   &$-$0.09   &$-$0.52   &$-$0.06   &   0.07   &$-$0.05   \\
Na (Na\,{\sc i})         &$-$0.22   &$-$0.11   &$-$0.20   &$-$0.36   &$-$0.04   &          &$-$0.61   &$-$0.24   &   0.22   &   0.07   \\
Mg (Mg\,{\sc i})         &$-$0.21   &$-$0.03   &$-$0.19   &$-$0.14   &$-$0.06   &          &$-$0.26   &$-$0.17   &   0.16   &$-$0.04   \\
Mg (MgH)                 &          &          &   0.11   &          &          &$-$0.43   &$-$0.05   &          &   0.09   &          \\
Al (Al\,{\sc i})         &$-$0.28   &$-$0.07   &$-$0.03   &$-$0.10   &$-$0.03   &          &          &$-$0.09   &   0.31   &$-$0.08   \\
Si (Si\,{\sc i})         &$-$0.12   &$-$0.02   &$-$0.12   &$-$0.11   &$-$0.01   &          &$-$0.33   &$-$0.06   &   0.40   &   0.08   \\
Si (Si\,{\sc ii})        &$-$0.17   &$-$0.06   &$-$0.21   &$-$0.33   &$-$0.19   &          &$-$0.54   &          &   0.59   &   0.11   \\
S (S\,{\sc i})           &          &$-$0.01   &          &$-$0.19   &          &          &$-$0.45   &          &   0.42   &$-$0.01   \\
K (K\,{\sc i})           &   0.08   &          &$-$0.26   &$-$0.32   &          &          &$-$0.41   &$-$0.20   &   0.01   &   0.33   \\
Ca (Ca\,{\sc i})         &$-$0.20   &$-$0.15   &$-$0.06   &$-$0.10   &$-$0.04   &          &$-$0.38   &$-$0.17   &   0.08   &   0.05   \\
Sc (Sc\,{\sc ii})        &$-$0.22   &$-$0.17   &$-$0.20   &$-$0.09   &$-$0.08   &   0.00   &$-$0.57   &$-$0.04   &   0.25   &$-$0.04   \\
Ti (Ti\,{\sc i})         &$-$0.07   &$-$0.21   &$-$0.12   &$-$0.08   &          &          &$-$0.37   &$-$0.16   &   0.10   &   0.13   \\
Ti (Ti\,{\sc ii})        &$-$0.34   &          &   0.06   &          &          &          &$-$0.40   &          &          &   0.03   \\
V (V\,{\sc i})           &          &          &$-$0.22   &$-$0.15   &          &          &          &$-$0.15   &   0.13   &          \\
Cr (Cr\,{\sc i})         &$-$0.33   &$-$0.13   &$-$0.36   &$-$0.40   &$-$0.06   &          &$-$0.67   &$-$0.22   &   0.08   &$-$0.03   \\
Cr (Cr\,{\sc ii})        &$-$0.26   &$-$0.08   &$-$0.21   &$-$0.22   &          &          &$-$0.57   &$-$0.03   &   0.38   &   0.05   \\
Fe (Fe\,{\sc i})         &$-$0.24   &$-$0.17   &$-$0.42   &$-$0.46   &$-$0.06   &$-$0.10   &$-$0.67   &$-$0.20   &   0.13   &   0.01   \\
Fe (Fe\,{\sc ii})        &$-$0.18   &$-$0.14   &$-$0.36   &$-$0.33   &$-$0.05   &          &$-$0.76   &$-$0.06   &   0.40   &   0.10   \\
Co (Co\,{\sc i})         &          &$-$0.10   &$-$0.32   &$-$0.30   &          &          &$-$0.62   &$-$0.11   &   0.21   &          \\
Ni (Ni\,{\sc i})         &$-$0.24   &$-$0.18   &$-$0.32   &$-$0.35   &$-$0.06   &$-$0.01   &$-$0.66   &$-$0.15   &   0.22   &   0.00   \\
Cu (Cu\,{\sc i})         &$-$0.35   &          &          &$-$0.20   &          &$-$0.40   &$-$0.80   &$-$0.20   &   0.10   &   0.00   \\
Zn (Zn\,{\sc i})         &          &$-$0.11   &          &$-$0.30   &          &   0.16   &$-$0.49   &          &   0.45   &$-$0.05   \\
Y (Y\,{\sc ii})          &$-$0.42   &          &$-$0.30   &          &          &          &          &          &          &          \\
Zr (Zr\,{\sc i})         &          &          &          &          &          &          &          &          &   0.03   &          \\
Nd (Nd\,{\sc ii})        &          &$-$0.22   &$-$0.17   &$-$0.17   &          &          &$-$0.59   &          &   0.33   &   0.02   \\
$\alpha$                 &$-$0.15   &$-$0.10   &$-$0.12   &$-$0.11   &$-$0.04   &          &$-$0.34   &$-$0.14   &   0.19   &   0.06   \\
\hline
\end{tabular}
\end{table*}

\begin{table*}
\setcounter{table}{5}
\caption{ Continued }
\begin{tabular}{l r r r r r r r r r r}
HD                       &   62644  &   82734  &  111028  &  115577  &  130952  &  142198  &  168723  &  196171 &  197964  &  199623  \\
\hline
Label                    &    SGB   &  SGB     & RGB      & RGB      & RGB      &  RGB     & RGB      & RGB     &  RGB     &    SGB   \\
\Teff $[$K$]$            &  5143    &  4990    &  4710    &  4754    &  4688    &  4800    &  4941    &  4697   &  4800    &  6246    \\
$\log g$                 &  3.42    &  3.11    &  3.10    &  2.75    &  2.43    &  2.91    &  2.95    &  3.00   &  3.25    &  4.27    \\
$[$M/H$]$                &$-$0.11   &   0.29   &$-$0.06   &$-$0.42   &$-$0.28   &$-$0.10   &$-$0.14   &$-$0.07  &   0.30   &$-$0.32   \\
$\xi_{\rm }t$ $[$km/s$]$ &  1.60    &  2.00    &  1.00    &  1.00    &  1.20    &  1.20    &  1.00    &  1.00   &  1.20    &  1.65    \\
M$_{\rm V}$              &  3.13    &  3.51    &  2.40    &  0.51    &  0.82    &  0.64    &  1.84    &  0.65   &  1.81    &  3.48    \\
Age $[$Gyr$]$            &  6       &  4.5     & 13       &  2       &  6       &  1       &  4       &  1      &  1       &  4       \\
Mass $[$M$_\odot$$]$     &   1.2    &   1.3    &   1.0    &   1.9    &   1.5    &   2.0    &   1.5    &   2.0   &  2.0     &   1.2    \\
$U$ $[$km/s$]$           &    56    &  $-$5    &   116    &    59    &    99    &    15    &    52    &     6   &    28    &     4    \\
$V$ $[$km/s$]$           & $-$25    &  $-$4    & $-$52    &$-$166    & $-$12    &    44    & $-$61    &    15   & $-$16    &    26    \\
$W$ $[$km/s$]$           & $-$35    &    10    &    29    &    39    &    39    &    14    &    22    &     2   &  $-$4    &    30    \\
$V_{\rm tot}$ $[$km/s$]$ &    71    &    12    &   131    &   180    &   107    &    49    &    84    &    17   &    32    &    40    \\
Li (Li\,{\sc i})         &$\le-$0.68&   0.21   &          &$\le-$1.31&$\le-$1.21&$\le-$1.06&$\le-$0.88&         &$\le-$0.59&   1.39   \\
C ($[$C\,{\sc i}$]$)     &$-$0.21   &   0.11   &$-$0.52   &$-$0.62   &$-$0.31   &$-$0.22   &$-$0.47   &$-$0.15  &          &$-$0.12   \\
C (CH)                   &$-$0.27   &          &          &$-$0.53   &$-$0.45   &$-$0.22   &          &         &          &          \\
C (C$_2$)                &$-$0.22   &          &          &$-$0.50   &$-$0.45   &$-$0.26   &$-$0.43   &$-$0.25  &   0.01   &          \\
$^{12}$C/$^{13}$C        &          &          &          &          &          & 12$\pm$3 & 40$\pm$25& 18$\pm$6& 40$\pm$15&          \\
N (N\,{\sc i})           &          &          &          &          &          &          &          &         &          &          \\
N (CN)                   &$-$0.28   &          &          &$-$0.40   &$-$0.25   &   0.04   &$-$0.02   &   0.24  &   0.36   &          \\
O ($[$O\,{\sc i}$]$)     &$-$0.16   &   0.25   &$-$0.09   &$-$0.17   &$-$0.36   &$-$0.17   &$-$0.32   &$-$0.09  &   0.24   &$-$0.03   \\
Na (Na\,{\sc i})         &$-$0.09   &   0.65   &   0.00   &$-$0.32   &$-$0.23   &$-$0.02   &$-$0.06   &$-$0.06  &   0.36   &$-$0.37   \\
Mg (Mg\,{\sc i})         &$-$0.04   &   0.23   &   0.04   &$-$0.06   &   0.02   &$-$0.06   &$-$0.10   &$-$0.06  &   0.25   &$-$0.23   \\
Mg (MgH)                 &$-$0.20   &          &          &$-$0.13   &$-$0.16   &$-$0.20   &$-$0.22   &   0.18  &   0.18   &          \\
Al (Al\,{\sc i})         &$-$0.01   &   0.32   &          &   0.02   &   0.00   &   0.04   &$-$0.01   &         &   0.38   &$-$0.25   \\
Si (Si\,{\sc i})         &   0.06   &   0.52   &   0.21   &   0.02   &   0.15   &   0.14   &   0.01   &   0.26  &   0.47   &$-$0.23   \\
Si (Si\,{\sc ii})        &   0.23   &   0.76   &   0.25   &$-$0.02   &   0.18   &   0.34   &$-$0.07   &   0.34  &   0.62   &          \\
S (S\,{\sc i})           &   0.04   &   0.59   &   0.18   &$-$0.35   &   0.18   &   0.17   &   0.02   &   0.44  &          &$-$0.29   \\
K (K\,{\sc i})           &$-$0.02   &          &$-$0.09   &          &   0.08   &   0.06   &   0.15   &         &   0.19   &   0.00   \\
Ca (Ca\,{\sc i})         &$-$0.08   &   0.42   &   0.11   &$-$0.25   &$-$0.01   &   0.00   &   0.01   &   0.00  &   0.28   &$-$0.24   \\
Sc (Sc\,{\sc ii})        &$-$0.11   &   0.45   &   0.08   &$-$0.15   &$-$0.13   &   0.08   &$-$0.16   &   0.01  &   0.35   &$-$0.29   \\
Ti (Ti\,{\sc i})         &$-$0.25   &   0.31   &$-$0.05   &$-$0.17   &$-$0.20   &   0.04   &$-$0.02   &$-$0.24  &   0.25   &          \\
Ti (Ti\,{\sc ii})        &          &          &          &          &          &          &          &         &          &          \\
V (V\,{\sc i})           &$-$0.39   &   0.30   &   0.00   &$-$0.24   &          &          &          &         &   0.35   &          \\
Cr (Cr\,{\sc i})         &$-$0.17   &          &$-$0.06   &$-$0.46   &$-$0.31   &          &$-$0.14   &         &   0.18   &$-$0.25   \\
Cr (Cr\,{\sc ii})        &$-$0.04   &          &   0.12   &$-$0.21   &          &          &   0.01   &         &          &$-$0.21   \\
Fe (Fe\,{\sc i})         &$-$0.11   &   0.29   &$-$0.06   &$-$0.42   &$-$0.28   &$-$0.10   &$-$0.14   &$-$0.07  &   0.30   &$-$0.32   \\
Fe (Fe\,{\sc ii})        &$-$0.03   &   0.61   &   0.18   &$-$0.26   &   0.10   &   0.20   &$-$0.03   &   0.30  &   0.49   &$-$0.19   \\
Co (Co\,{\sc i})         &$-$0.25   &          &$-$0.10   &$-$0.42   &$-$0.19   &   0.00   &$-$0.21   &         &   0.33   &          \\
Ni (Ni\,{\sc i})         &$-$0.11   &   0.41   &   0.04   &$-$0.34   &$-$0.24   &   0.02   &$-$0.07   &$-$0.04  &   0.42   &$-$0.36   \\
Cu (Cu\,{\sc i})         &          &          &          &          &          &          &          &         &          &          \\
Zn (Zn\,{\sc i})         &   0.06   &          &   0.19   &   0.05   &   0.21   &   0.42   &$-$0.03   &   0.19  &          &          \\
Y (Y\,{\sc ii})          &          &          &          &          &          &          &          &         &          &          \\
Zr (Zr\,{\sc i})         &          &   0.49   &          &          &          &          &          &$-$0.26  &   0.20   &          \\
Nd (Nd\,{\sc ii})        &$-$0.19   &          &   0.04   &$-$0.33   &$-$0.01   &          &   0.03   &         &          &          \\
$\alpha$                 &$-$0.08   &   0.37   &   0.09   &$-$0.11   &$-$0.01   &   0.03   &$-$0.03   &$-$0.11  &   0.31   &$-$0.23   \\
\hline
\end{tabular}
\end{table*}

\begin{table}
\setcounter{table}{5}
\caption{ Continued }
\begin{tabular}{l r r r}
HD                       &  207978  &  212487  &  219617 \\
\hline
Label                    &  SGB     & SGB      & SGB     \\
\Teff $[$K$]$            &  6400    &  6202    &  5762   \\
$\log g$                 &  3.90    &  3.60    &  3.90   \\
$[$M/H$]$                &$-$0.47   &$-$0.14   &$-$1.55  \\
$\xi_{\rm t}$ $[$km/s$]$ &  1.30    &  1.80    &  2.00   \\
M$_{\rm V}$              &  3.31    &  2.59    &  3.56   \\
Age $[$Gyr$]$            &  4       &  3       & 13      \\
Mass $[$M$_\odot$$]$     &   1.2    &   1.4    &   0.8   \\
$U$ $[$km/s$]$           &    24    &    60    &   392   \\
$V$ $[$km/s$]$           &    21    &     3    &$-$326   \\
$W$ $[$km/s$]$           &     0    &  $-$3    & $-$48   \\
$V_{\rm tot}$ $[$km/s$]$ &    32    &    60    &   512   \\
Li (Li\,{\sc i})         &$\le$0.44 &   1.11   &   0.91  \\
C ($[$C\,{\sc i}$]$)     &          &          &         \\
C (CH)                   &$-$0.64   &$-$0.20   &         \\
C (C$_2$)                &          &          &         \\
$^{12}$C/$^{13}$C        &          &          &         \\
N (N\,{\sc i})           &          &          &         \\
N (CN)                   &          &          &         \\
O ($[$O\,{\sc i}$]$)     &$-$0.33   &$-$0.21   &$-$0.88  \\
Na (Na\,{\sc i})         &$-$0.42   &$-$0.14   &$-$1.64  \\
Mg (Mg\,{\sc i})         &$-$0.48   &$-$0.22   &$-$1.28  \\
Mg (MgH)                 &          &          &         \\
Al (Al\,{\sc i})         &$-$0.50   &$-$0.14   &$-$1.28  \\
Si (Si\,{\sc i})         &$-$0.33   &$-$0.14   &         \\
Si (Si\,{\sc ii})        &$-$0.49   &          &         \\
S (S\,{\sc i})           &          &   0.06   &         \\
K (K\,{\sc i})           &$-$0.23   &   0.27   &$-$1.17  \\
Ca (Ca\,{\sc i})         &$-$0.41   &$-$0.10   &$-$1.26  \\
Sc (Sc\,{\sc ii})        &$-$0.40   &$-$0.30   &         \\
Ti (Ti\,{\sc i})         &          &$-$0.03   &         \\
Ti (Ti\,{\sc ii})        &$-$0.45   &$-$0.31   &$-$1.45  \\
V (V\,{\sc i})           &          &          &         \\
Cr (Cr\,{\sc i})         &          &          &         \\
Cr (Cr\,{\sc ii})        &          &          &         \\
Fe (Fe\,{\sc i})         &$-$0.47   &$-$0.14   &$-$1.55  \\
Fe (Fe\,{\sc ii})        &$-$0.50   &$-$0.14   &$-$1.55  \\
Co (Co\,{\sc i})         &          &          &         \\
Ni (Ni\,{\sc i})         &$-$0.51   &$-$0.20   &$-$1.58  \\
Cu (Cu\,{\sc i})         &          &          &         \\
Zn (Zn\,{\sc i})         &          &          &         \\
Y (Y\,{\sc ii})          &$-$0.66   &$-$0.40   &$-$1.93  \\
Zr (Zr\,{\sc i})         &          &          &         \\
Nd (Nd\,{\sc ii})        &          &          &         \\
$\alpha$                 &$-$0.41   &$-$0.12   &$-$1.27  \\
\hline
\end{tabular}
\end{table}

\begin{figure*}[ht]
 \resizebox{\hsize}{!}{\includegraphics{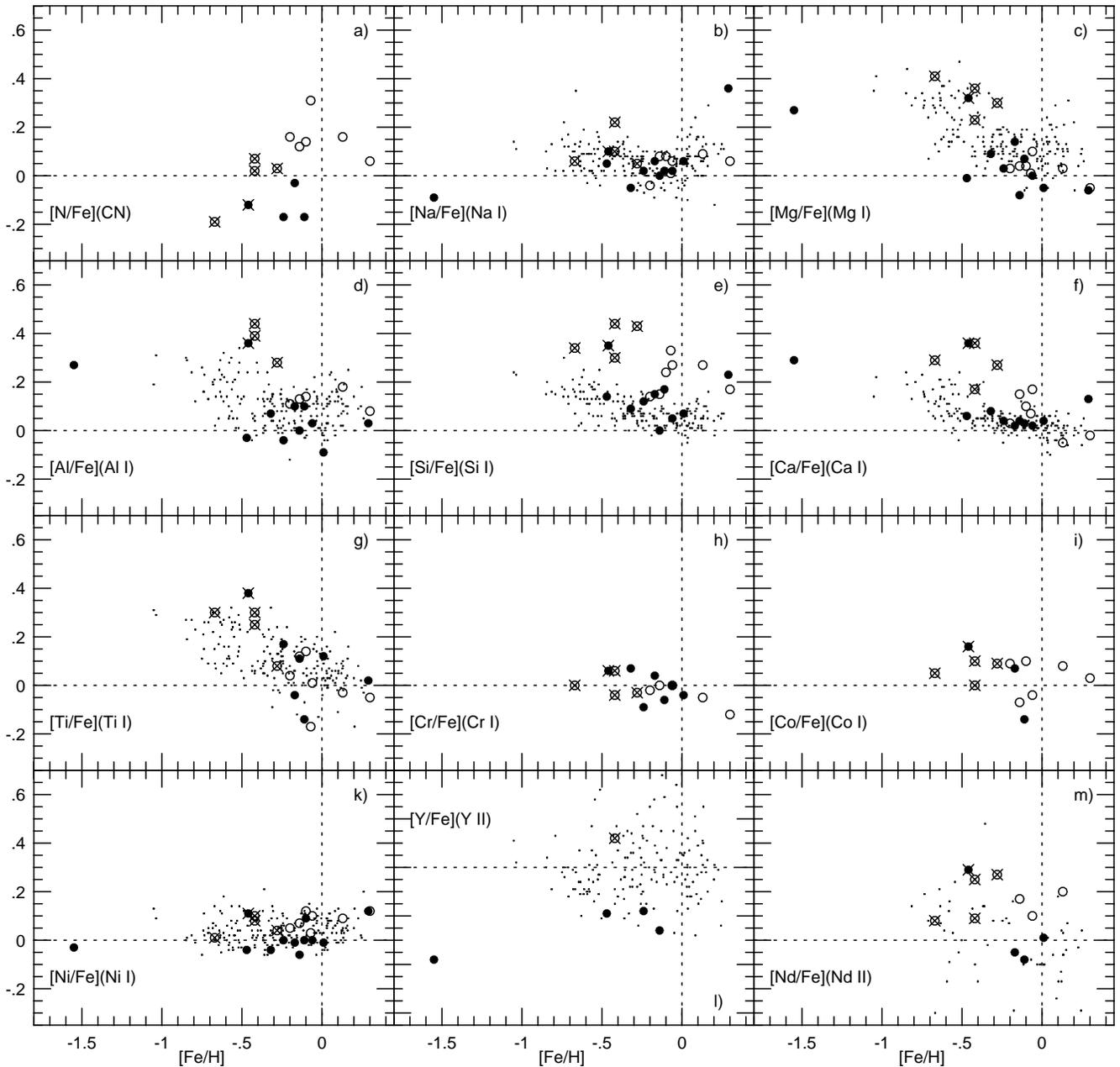}}
  \caption{Abundance-abundance diagrams for several element ratios.
   SGB stars are shown as solid dots while RGB stars are circles.
   Five candidate thick-disk stars are identified by crossed symbols.
   The small dots are disk star abundances from EAGLNT.
   The vertical scale for yttrium has been shifted by 0.3\,dex }
 \label{allab}
\end{figure*}

The resulting chemical abundances relative to the Sun
are presented in Table\,\ref{abund1} and shown in several figures below.
In the following section patterns in the element ratios are discussed.
Fig.\,\ref{allab} displays element [X/Fe] vs. [Fe/H] (derived from lines of
Fe\,{\sc i}) for several species.
The abundances derived for disk stars by EAGLNT
are shown as small dots for comparison.

In Figures \ref{allab} to \ref{sifeteff} the subgiant branch (SGB) stars are
represented by solid dots, while the results for stars which have started to
ascend the giant branch (RGB) are shown as open circles.
Note that the RGBs sometimes display a different abundance pattern as compared
to the SGBs which are the primary targets for this study.
The SGB star HD\,18907 and the four RGB stars HD\,6734, HD\,24616, HD\,115577
and HD\,130952 are believed to be thick-disk stars and are identified by crossed
symbols in the figures.
These were identified in Sect.\,\ref{dynamics} and are discussed separately in
Sect.\,\ref{thickdisk}.

\subsection{Lithium}

Li is produced in the Big Bang and later by
cosmic ray spallation and in different stellar objects
(e.g. Guessoum \& Kazanas 1999; Denissenkov \& Weiss 2000;
Ramaty et al. 2000; Travaglio et al. 2001).

\begin{figure*}[ht]
 \resizebox{\hsize}{!}{\includegraphics{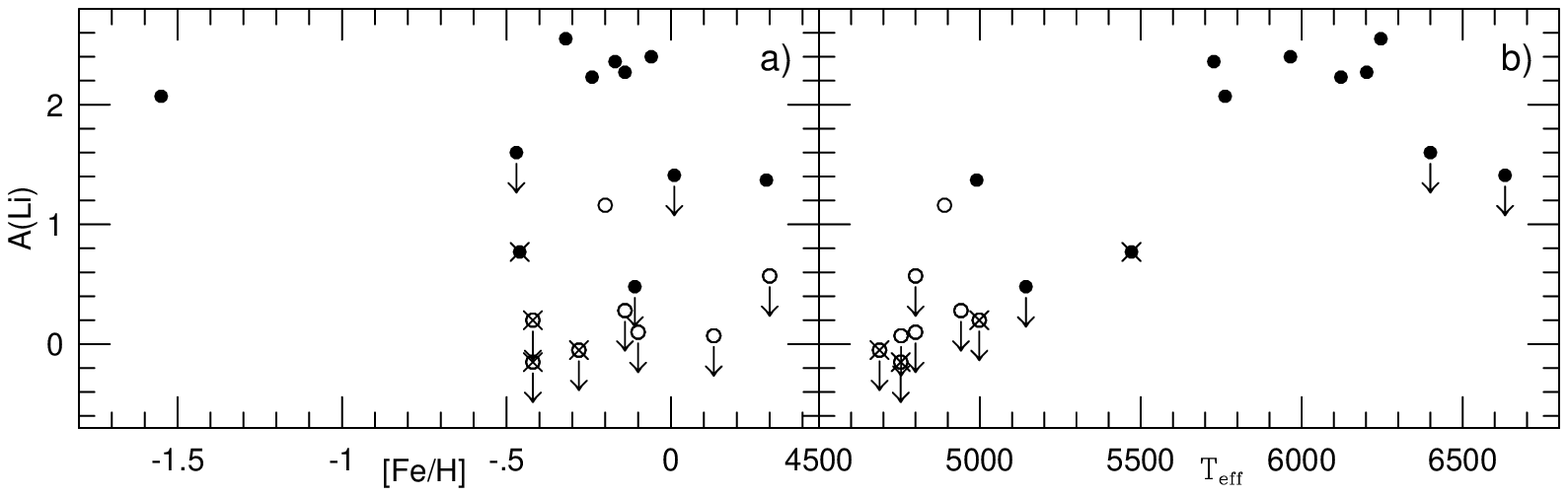}}
  \caption{Li abundances vs. metallicity and effective temperature.
   Symbols as in Fig.\,\ref{allab}, and the arrows denote upper limits }
 \label{liloggteff}
\end{figure*}

Li is destroyed by proton capture at low stellar interior temperatures,
leading to Li depletion in stellar atmospheres for which the surface
matter has been subject to convective mixing with deeper regions
(cf., e.g., Boesgaard 1991; Deliyannis et al. 1998).
We used the 6707\,\AA\ doublet to derive the total Li abundance
($^6$Li\,+\,$^7$Li).
Li region spectra were secured for 21 stars.
The line was detected in 11 stars while for 10 stars only an upper limit could
be obtained.
Fig.\,\ref{liloggteff}a displays a large spread in the Li abundance diagram.
Several disk stars show only low upper limits, consistent with evolutionary
depletion from an initial abundance of A(Li) above 3.0.
Only five disk-SGB stars show abundances above
A(Li)\,$=\log (N_{\rm Li}/N_{\rm H})+12=2.0$.
These are only slightly metal deficient and seem to be depleted
by about 1\,dex relative to the upper envelope of A(Li) vs. [Fe/H]
distribution as compiled e.g. in Fig.\,6 of Ryan et al. (2001).
Our halo SGB star, HD\,219617 with [Fe/H]\,$=-1.55$, seems to be typical
for or only slightly below the locus for non-depleted halo dwarfs.

The observed pattern also shows a \Teff dependent effect.
Cool main sequence stars are well known to destroy Li,
and the deepening surface convection zone as the stars evolve towards the giant
branch will cause further Li destruction.
Fig.\,\ref{liloggteff}b shows that the stars with
detected Li divides into two groups: The stars with
$T_{\rm eff} < 5500$ have
Li abundances A(Li)\,$\leq 1.5$.
The warmer group of stars have higher abundances but still (with the exception
of HD\,219617) lower than their presumed primordial abundances.
Our two warmest stars,  the subgiants Procyon (HD\,61421) and HD\,207987 show
no traces of the Li doublet.
These may already on the main sequence have destroyed most of their Li
(Boesgaard \& Trippicco 1986).
A plot of Li abundance vs. stellar mass, tracing main sequence $T_{\rm eff}$,
does not show any evident correlation.
An elaborate discussion of Li based on a large sample of subgiant stars was
recently given by do Nascimento et al. (2003)\nocite{nascimento}.

\subsection{Carbon}

Traces of post-main-sequence evolution could be expected for C,
since this element is converted to nitrogen in the CN cycle,
which is weak but existing in solar-mass dwarfs
(Bahcall \& Ulrich 1988)\nocite{bahcall}, and which dominates the energy
generation during the shell-hydrogen-burning phase.
The effects of CN cycling is seen in the surface abundances of stars which
have experienced the first dredge up.

\begin{figure}[ht]
 \resizebox{\hsize}{!}{\includegraphics{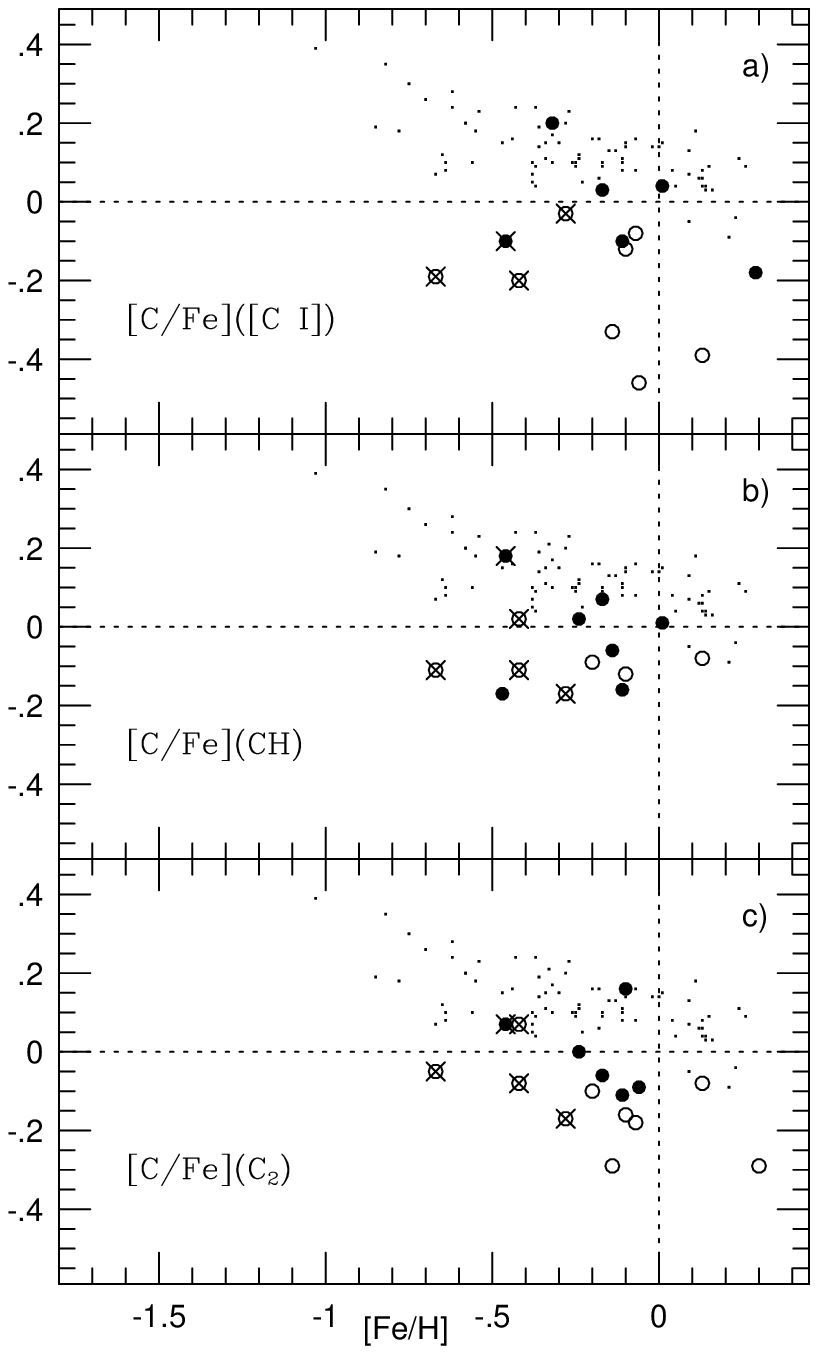}}
  \caption{ [C/Fe] from [C\,{\sc i}], CH and C$_2$ lines.
   Symbols as in Fig.\,\ref{allab} }
 \label{cfeall}
\end{figure}

\begin{figure}[ht]
 \resizebox{\hsize}{!}{\includegraphics{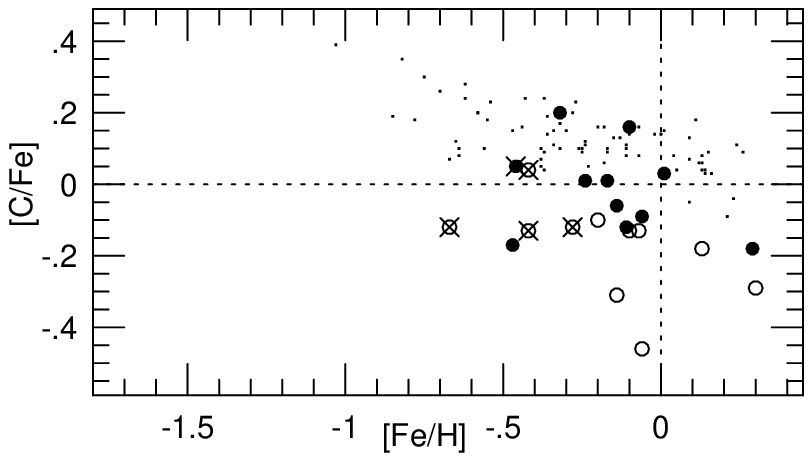}}
  \caption{ Mean value of [C/Fe] from [C\,{\sc i}], CH and C$_2$ lines.
   Symbols as in Fig.\,\ref{allab} }
 \label{cfe}
\end{figure}

The weak, forbidden [C\,{\sc i}] line at 8727\,\AA\ was used for C abundance
derivation for 80 F--G dwarfs by Gustafsson et al. (1999).
They found a weakly increasing [C/Fe] trend with decreasing metallicity.
This data are shown as small dots in Figs.\,\ref{cfeall} and \ref{cfe}.
Here, the same forbidden line with a complementary highly excited C\,{\sc i}
line (7662\,\AA, in the middle of a strong telluric molecular line forest and
only visible in a few of our warmest stars) could be used for half our stars.
Our subgiants (solid dots) in Fig.\,\ref{cfeall}a scatter more and show a
somewhat lower mean value than do the dwarfs of Gustafsson et al.
Only 0.03\,dex of the difference can be blamed on the difference in $gf$ values.
The first line of Table\,\ref{sgbscatter} (in Sect.\,\ref{discuss})
compares the scatter relative to
linear regressions of [C\,{\sc i}]-line-based [C/Fe] vs. [Fe/H] for our SGBs
(0.09\,dex) and the dwarfs of Gustafsson et al. (0.06\,dex).

The CH- and C$_2$-line results are shown in Fig.\,\ref{cfeall}b) and c).
All three criteria give similar results with giants generally showing lower
carbon abundances than subgiant stars at similar metallicity.
Five of the subgiants seem, however, to have low carbon abundances similar
to those of the least depleted giants in Fig.\,\ref{cfe},
where the straight-mean values of all carbon-abundance criteria are shown.
This may possibly be an indication of deep mixing in stars which
have not yet reached the giant branch.
We do, however, not find any correlation between the subgiant [C/Fe] and
either $M_{\rm V}$, stellar mass, or Li abundance.

\begin{figure}[ht]
 \resizebox{\hsize}{!}{\includegraphics{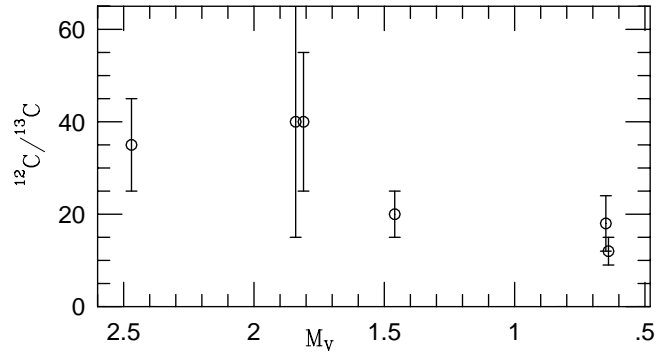}}
  \caption{$^{12}$C/$^{13}$C vs. absolute magnitude
   Symbols as in Fig.\,\ref{allab} }
 \label{C13_Mv}
\end{figure}

$^{12}$C/$^{13}$C abundance ratios could be derived for six of the giant
stars with roughly solar metallicity.
For the remaining stars the CN (A-X) lines were too week or not observed.
In Fig.\,\ref{C13_Mv} the derived ratios are plotted vs. absolute magnitude.
The theoretical isotopic ratios expected after the first dredge-up for stars
of this metallicity and mass range are in the range 20-30
(Boothroyd \& Sackmann 1999, Table\,2; Girardi et al. 2000;
Salaris et al. 2002).
The three most luminous of the six, $M_V<1.5$, also show the lowest carbon
isotope ratios.
$^{12}$C/$^{13}$C\,=\,$12 \pm 3$ for HD\,142198 requires
models with extra mixing after the first dredge-up for its explanation
(see e.g., the recipe for ``Cool Bottom Processing'' in Table\,2 of
Boothroyd \& Sackmann 1999).

\subsection{Nitrogen}

\begin{figure}[ht]
 \resizebox{\hsize}{!}{\includegraphics{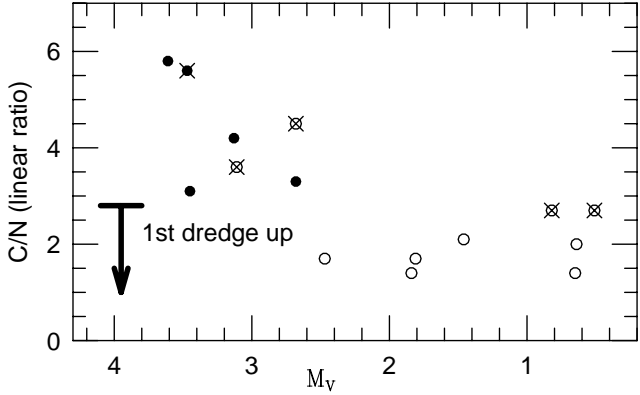}}
  \caption{ $N_{\rm C}/N_{\rm N}$ vs. absolute magnitude.
    Symbols as in Fig.\,\ref{allab}. The arrow shows the predicted
    effect of the 1st dredge-up according to
    Boothroyd \& Sackmann (1999) }
 \label{cnmv}
\end{figure}

The same mechanism which depletes carbon on the giant branch would also
bring fresh nitrogen to the surface.
This is probably seen in Fig.\,\ref{allab}a).
Our N\,{\sc i} line could only be measured in two stars, while the CN
molecular lines could be used for half of the sample.
For the subgiant HD\,2151, the only star for which both criteria could be used,
the N\,{\sc i} and CN criteria give different abundances by a factor of two
(0.3\,dex), which calls for caution in the interpretation.
One should bear in mind that the N abundances derived from CN lines are quite
sensitive to the adopted carbon abundances.
The linear abundance ratios $N_{\rm C}/N_{\rm N}$, using the mean values of
our C and N abundance criteria,
are plotted against absolute magnitude in Fig.\,\ref{cnmv}.
Consider separately the five crossed symbols, representing thick-disk stars,
which may have had particular initial abundance ratios as discussed below in
Sects.\,\ref{alpha} and \ref{thickdisk} and which
also have lower metallicities than the other objects in this figure.
This quintuplet may describe a parallel trend to the ``normal'' disk stars,
which are discussed first:
note that the six stars with $N_{\rm C}/N_{\rm N}\le 2.1$
in Fig.\,\ref{cnmv} are exactly the same giants for which we could estimate
$^{12}$C/$^{13}$C ratios and thus verify surface abundance changes:
HD\,35410, HD\,40409, HD\,142198, HD\,168723, HD\,196171 and HD\,197964.
Boothroyd \& Sackmann (1999, Table\,2, BS99 below) tabulate the expected
surface abundances after the 1st dredge up for stars of different masses.
Their initial C/N number abundance ratio before dredge-up is however only 2.8
and their C/N given after the first dredge up for stars with masses
in the range 1.3--2.3\,$M_\odot$ fall between 1.2 and 0.8, respectively.
This is indicated by the thick arrow.
Our pre-dredge-up C/N abundance ratio (as judged from the four thin-disk
subgiants in Fig.\,\ref{cnmv}) is almost two times higher than that assumed
by BS99.
If a simple scaling of the BS99 first dredge-up C/N ratios by the initial C/N
is roughly permissible, also our C/N ratios agree well with 1st dredge-up
predictions, which are not noticeably affected by possible
``Cool Bottom Processing''.

\begin{figure}[ht]
 \resizebox{\hsize}{!}{\includegraphics{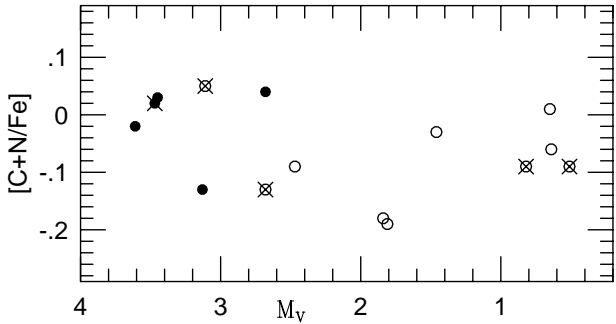}}
  \caption{The concentration of C+N nuclei relative to iron is independent of
   absolute magnitude as expected from the processes of CN cycling.
   Symbols as in Fig.\,\ref{allab} }
 \label{candnmv}
\end{figure}

The first dredge-up episode should show the effects of the conversion of carbon
to nitrogen, while preserving the sum of the two nuclei.
Fig.\,\ref{candnmv} shows that [C+N/Fe], i.e. the sum of C and N
normalized to the metallicity,
does not scatter by more than 0.08\,dex (20\%) relative to a
mean value of $-0.06$, and that there is no significant trend with luminosity.

\subsection{Oxygen}
\label{oxygen}

\begin{figure}[ht]
 \resizebox{\hsize}{!}{\includegraphics{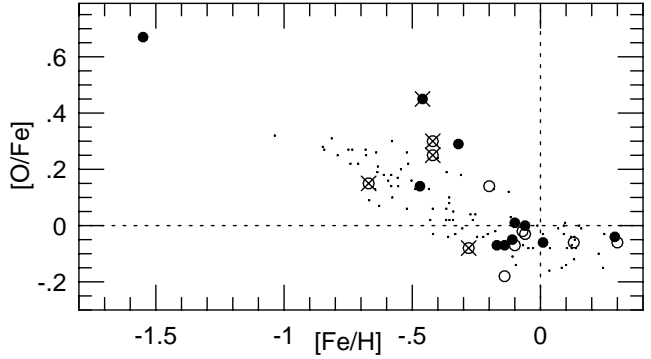}}
  \caption{[O/Fe] vs. [Fe/H]. Symbols as in Fig.\,\ref{allab} }
 \label{oxfig}
\end{figure}

ISM oxygen is supposedly produced almost exclusively in stars ending their
lifes as core-collapse supernovae.
The weak forbidden [O\,{\sc i}] lines at 6300 and 6363\,\AA\ were used to
estimate oxygen abundances, as well as highly excited lines in the 6150\,\AA\
region, the latter usable only for a few hot stars.
The 6363\,\AA\ forbidden line is in the wings of a strong Ca\,{\sc i}
autoionisation line at 6361.8\,\AA, which was simulated by application of an
artificially magnified radiation damping parameter,
$\Gamma_{\rm rad} = 10^{12} s^{-1}$.
In all stars except HD\,400 one or both of the forbidden lines have been
detected.
[O/Fe] increases with decreasing [Fe/H] which is shown in Fig.\,\ref{oxfig}.
The subgiants display a smaller scatter around solar metallicity in
Fig.\,\ref{oxfig} than do the dwarfs of EAGLNT, while they may possibly be
more oxygen rich than EAGLNT around [Fe/H]$\approx -0.4$.

\subsection{The $\alpha$ elements}
\label{alpha}

\begin{figure*}[ht]
 \resizebox{\hsize}{!}{\includegraphics{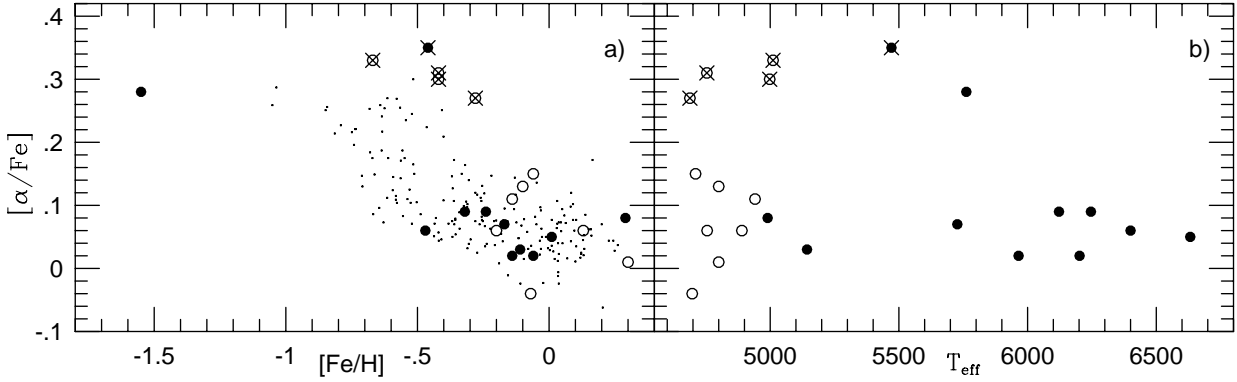}}
  \caption{[$\alpha$/Fe] vs. metallicity and effective temperature.
   For the thin-disk stars there is no trend of [$\alpha$/Fe]
   with effective temperature.  Symbols as in Fig.\,\ref{allab} }
 \label{alphafeteff}
\end{figure*}

Mg, Si, S and Ca, our $\alpha$ elements, are like oxygen believed to be
predominantly formed in massive stars and may therefore be expected to
show similar abundance patterns. Mg is suspected to have an origin more in
common with O. Titanium is sometimes expected to be formed as an iron peak
element but usually shows abundance patterns similar to the $\alpha$
elements, c.f., e.g. EAGLNT. This is also a result of this investigation,
why we include it among the $\alpha$s. The [Mg/H] values derived from MgH
lines from 9 giants and 2 subgiants agree with the the Mg\,{\sc i}-line
results with a scatter of 0.17\,dex, but they show a steeper trend with
metallicity (not plotted). In the panels for [Si/Fe] and [Ca/Fe],
Fig.\,\ref{allab}e) and f), several giants show abundances which are
higher than those of the subgiants. The reason for this behaviour is not
known, and it remains after a simple modification of e.g. the effective
temperatures. We are not aware of any nucleosynthesis process which could
alter these $\alpha$-element abundances during early post-main-sequence
evolution. We note that the subgiants also tend to follow the trends found
for disk dwarfs, as we would by default expect for all our stars. It can
not be excluded that the more deviating trends found for the giants
disclose problems in the analysis. One possibility could be a
differentially stronger overionization in Fe than in Mg, Si and Ca (cf.
Fig.\,\ref{overion}) leading to overestimated [Mg,Si,Ca/Fe]. The sulfur
abundance is derived from two highly excited S\,{\sc i} lines and [S/Fe]
only shows a large scatter around a mean of $+0.20$ (not plotted).

\begin{figure}[b]
 \resizebox{\hsize}{!}{\includegraphics{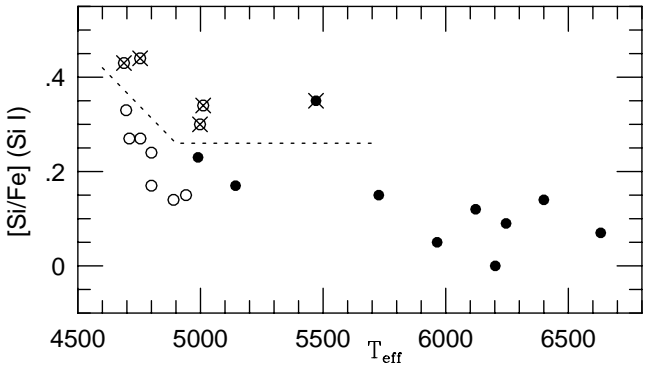}}
  \caption{
   Thin-disk stars, below the dotted line, show a systematic Si abundance
   trend with \Teff. Such a trend is absent for the mean $\alpha$ abundance
   in Fig.\,\ref{alphafeteff}b). Symbols as in Fig.\,\ref{allab} }
 \label{sifeteff}
\end{figure}

Fig.\,\ref{alphafeteff}a) shows [$\alpha$/Fe] vs. [Fe/H], where
[$\alpha$/Fe]\,$= {1\over 4}($[Mg/Fe]$+$[Si/Fe]$+$[Ca/Fe]$+$[Ti/Fe]$)$ as
computed from the results from neutral atomic lines is defined as in EAGLNT.
Panel b) shows [$\alpha$/Fe] vs. $T_{\rm eff}$.
The disk-metallicity stars, excluding the group of five with crossed symbols,
show no trend of [$\alpha$/Fe] with effective temperature.
Such trends, with different signs, do show up in individual
abundance ratios relative to Fe, see e.g. Fig.\,\ref{sifeteff}, but each
of them is based on fewer spectral lines.
We consider our mean [$\alpha$/Fe] ratios to be more reliable
than those of individual $\alpha$ elements.

Most of our subgiant stars show mean $\alpha$ element abundances in excellent
agreement with the disk dwarf stars of EAGLNT in Fig.\,\ref{alphafeteff}a.
A group of five stars, with overcrossed symbols in most of our figures and
including the subgiant HD\,18907, shows ratios of [Al/Fe], [Si/Fe] and [Ca/Fe],
which deviate markedly from those
of our other disk stars and from EAGLNT stars at similar metallicities.
Also for [Mg/Fe] and [Ti/Fe] the group falls above the mean trend of
the disk stars when plotted vs. [Fe/H], see Fig.\,\ref{allab}c) and g).
In general this group shows abundance ratios which are similar to those of
metal-poor stars, but for oxygen this is only true for the subgiant HD\,18907.
These ``chemical'' characteristics strengthen our proposal in
Sect.\,\ref{dynamics} that these stars belong to the thick disk of the Galaxy.
The [Fe/H]\,$=-0.06$ giant HD\,111028 has a rather large total space velocity
(131\,km/s) and also a relatively high $\alpha$-element abundance.
It may therefore be a metal-rich thick-disk star of the kind
recently discovered by Feltzing et al. (2003).

\subsection{Sodium, aluminium and potassium}

For Na and Al, Fig.\,\ref{allab}b) and d), the subgiants seem to follow the
trends outlined by the disk dwarfs, while $\alpha$-element-rich
thick-disk-candidate stars seem to be enriched also in Al.
Potassium was measured only from the outer parts of the strong wings
of two resonance lines and this data is therefore uncertain and
not plotted here.

\subsection{Iron peak elements}

The elements Cr and Ni follow Fe closely in Fig.\,\ref{allab}h) and
\ref{allab}k).
Cobalt in Fig.\,\ref{allab}i) also follows Fe but with a slightly larger
scatter.
The Co\,{\sc i} lines are weak (less than 10\,m\AA\ in the Sun) and hyperfine
structure was ignored in the analysis.
[V/Fe] (not plotted) is sensitive to effective temperature and hyperfine
structure, and the three thick-disk candidates with measured V\,{\sc i} lines
show ratios enhanced by about 0.25\,dex while one thin-disk subgiant
(HD\,62644) has [V/Fe]\,$=-0.28$).

The nucleosynthetic origins of copper and zinc are not well determined.
[Cu/Fe] generally scatters around zero, while [Zn/Fe] displays a scatter of
$\pm 0.2$\,dex around $+0.3$.
These are not plotted.

\subsection{s-process elements}

The elements tracing the s-process are only poorly examined in this study.
One Y\,{\sc i} line could be measured at low metallicities but was difficult
to measure in the metal-rich stars.
The Y abundance seems to be low in our subgiant stars, Fig.\,\ref{allab}l),
but the statistics are very poor.
Nd was only measurable in disk stars and by a single Nd\,{\sc ii} line,
and shows a similar pattern as the dwarfs of EAGLNT, see Fig.\,\ref{allab}m).

\section{Discussion}
\label{discuss}

One key question of this study is whether abundances derived from subgiant stars
are as reliable probes of the star-forming ISM as those derived from dwarfs.
In Table\,\ref{sgbscatter} we try to quantify the scatter in the abundance
ratios [X/Fe] for the SGB stars with [Fe/H]\,$>-1.0$ relative to a linear
fit in the [X/Fe] vs. [Fe/H] diagram and we compare it to the corresponding
quantities found for disk dwarf stars by EAGLNT and Gustafsson et al. (1999).
These studies are, in terms of spectral resolution, S/N and spectral coverage
roughly equivalent to the present one.
When inspecting Table\,\ref{sgbscatter} and the diagrams we find that the
scatter for the subgiants is indeed somewhat larger.
Most of this scatter is, however, due to two stars which seem to depart from
the others.
The first one is HD\,18907 at [Fe/H]\,$=-0.46$, which shows typical thick-disk
characteristics and abundances very similar to our four thick-disk giants,
c.f. Sect.\,\ref{thickdisk}.
The other deviating subgiant is HD\,82734 at [Fe/H]\,$=+0.29$, which is more
metal rich than any of the EAGLNT disk dwarfs that we use for comparison,
and which appears particularly enriched in Na, Si, Ca and possibly Ni.
If these stars are excluded one finds that the scatter is not larger than
expected from known errors, and in most cases not larger than that obtained
for dwarf stars in EAGLNT.
In fact: for half of the abundance ratios it is smaller than for the
dwarfs.

\begin{table*}\caption{\label{sgbscatter}
Comparison of the abundance scatter for our SGB stars with that found for disk
dwarf stars by EAGLNT and Gustafsson et al. (1999).
The scatter, $\sigma$, is the standard deviation from a linear fit where
[X/Fe]\,$= a + b$\,[Fe/H].  $n_*$\,= number of stars in the fit,
and $n_{\rm lines}$\,= maximum number of lines.
For consistency, carbon abundances from molecular lines are neglected for SGB
stars. Columns 5 and 6 are for SGB stars excluding HD\,18907 and HD\,82734,
see the text for an explanation}
\begin{tabular}{l r r r c r r c c r r r}
\noalign{\smallskip}
Abundance ratio        &\multicolumn{5}{c}{SGB stars} & & & &\multicolumn{3}{c}{Disk dwarfs} \\
                        \cline{2-8}                      \cline{10-12}
                       & & & & \multicolumn{4}{c}{Excluding 2 stars} \\
                                   \cline{5-8}
(species)              & $\sigma$ & $n_*$ & $n_{\rm lines}$ &
                       & $\sigma$ & $n_*$ & &
                       & $\sigma$ & $n_*$ & $n_{\rm lines}$ \\
\hline
{[C/Fe]} ([C\,{\sc i}])  & 0.12 &   6 &  1 & & 0.09 & 4 & & & 0.06 &  80 &  1 \\
{[O/Fe]} ([O\,{\sc i}])  & 0.13 &  10 &  4 & & 0.09 & 8 & & & 0.07 &  87 &  4 \\
{[Na/Fe]} (Na\,{\sc i})  & 0.09 &  10 &  4 & & 0.04 & 8 & & & 0.07 & 187 &  2 \\
{[Mg/Fe]} (Mg\,{\sc i})  & 0.10 &  10 &  6 & & 0.07 & 8 & & & 0.08 & 178 &  2 \\
{[Al/Fe]} (Al\,{\sc i})  & 0.12 &  10 &  2 & & 0.07 & 8 & & & 0.08 & 184 &  2 \\
{[Si/Fe]} (Si\,{\sc i})  & 0.10 &  10 &  9 & & 0.05 & 8 & & & 0.05 & 189 &  8 \\
{[Ca/Fe]} (Ca\,{\sc i})  & 0.10 &  10 &  4 & & 0.02 & 8 & & & 0.05 & 189 &  4 \\
{[Ti/Fe]} (Ti\,{\sc i})  & 0.14 &   7 &  6 & & 0.13 & 5 & & & 0.08 & 182 &  4 \\
{[Ni/Fe]} (Ni\,{\sc i})  & 0.06 &  11 & 10 & & 0.04 & 9 & & & 0.05 & 188 & 20 \\
{[Nd/Fe]} (Nd\,{\sc ii}) & 0.09 &   4 &  2 & & 0.03 & 3 & & & 0.16 &  52 &  2 \\
\hline
\end{tabular}
\end{table*}

\subsection{Element abundances affected by stellar evolution}

One very important issue concerns the extent to which mixing of internally
processed material to the surface may have affected the original
abundances from the time of formation of the present subgiant stars.
For lithium this is a ``problem'' already for main-sequence stars, and
we can probably see effects of depletion in all our giants and subgiants except
possibly in the subgiant HD\,199623.

The next group of elements to suspect are those taking part in the CN cycle.
Canonically, these effects should appear only after the first dredge-up
episode, when a star has already started to climb the giant branch.

$^{12}$C/$^{13}$C should show the first traces of CN cycling, but unfortunately
we could not measure this ratio in any subgiant star.
For the six giants where we could measure the ratio we find it to be lowered
relative to the ISM as might be expected.
For the giant stars we also see clear traces of carbon depletion in
Fig.\,\ref{cfe}.
Nine of the ten giants with nitrogen determinations show [N/Fe]\,$>0.0$, and
about 0.2\,dex higher values than typical for the subgiants.
This is another sign of CN cycling.
Figs.\,\ref{cnmv} and \ref{candnmv} show in an alternative way the effects of
CN cycling for the giant stars.

Six of the subgiants seem not to be depleted in carbon since they fall within
the scatter of the disk dwarf stars of Gustafsson et al. (1999).
The five remaining subgiants with carbon abundance determinations, however,
show [C/Fe] which are lower than those seen in disk dwarfs and quite similar to
those of our giants at similar [Fe/H].
Only one of these, HD\,62644, has a nitrogen determination which indicates a
linear C/N abundance ratio of 4.2, about two times higher than typical
for the CN-processed giants and typical for the four
non-carbon-depleted subgiants with nitrogen determinations.
A word of caution should be given regarding the nitrogen abundances mainly
derived from CN molecular lines, since these abundances are very sensitive to
the derived carbon abundances.

The carbon abundances derived here indicate, however, that subgiant stars
may already show signs of dredge-up of CN-processed material.
This should be investigated further.

For oxygen we can trace no differences in the [O/Fe] vs. [Fe/H] diagrams
when we compare with results for dwarf stars in Fig.\,\ref{oxfig}.
Thus, no traces of other proton capture processes, e.g. ON cycling,
are seen or expected.

\subsection{The candidate thick-disk stars}
\label{thickdisk}

In Sections\,\ref{dynamics} and \ref{alpha}
we have identified one subgiant, HD\,18907, and four giant stars
HD\,6734, HD\,24616, HD\,115577 and HD\,130952,
in the metallicity range $-0.67\le [$Fe/H$]\le -0.29$ which show
silicon, calcium and aluminium abundances which are significantly
higher than those of disk dwarf stars of similar metallicities in EAGLNT.
Also the magnesium and titanium abundances are higher than typical,
which makes the group very ``$\alpha$-element'' rich (Fig.\,\ref{allab}).
This derived abundance pattern is not likely to be the effect of
errors in the model parameter or deviations from LTE.
The abundance ratios relative to iron for this group of stars are
actually quite similar to those of halo-metallicity stars.

All of these chemically deviating stars have space velocities deviating
by more than 80\,km\,s$^{-1}$ from the LSR (Fig.\,\ref{vtotfe}),
and, except for one giant, they have $V$ velocities lagging
behind the orbital velocity of the LSR by more than 50\,km\,s$^{-1}$.
These are kinematic signatures which are typical of old stellar populations.

The kinematics, aluminium and $\alpha$-element enrichment suggest that the
group may belong to the so-called ``thick-disk'' population,
(Fuhrmann K. 1998\nocite{fuhrmann}; Gratton et al. 2000\nocite{gratton};
Prochaska et al. 2000\nocite{prochaska};
Pettinger et al. 2001\nocite{pettinger}; Bensby et al. 2003\nocite{bensby}),
which has recently been proposed to contain stars with metallicities even
higher than solar, see Feltzing et al. (2003).
We tentatively also find overabundances in Sc, V, and Zn relative to Fe;
thick-disk characteristics as proposed by
Prochaska et al.\nocite{prochaska}
The thick disk, however, is usually found to be older than about 10\,Gyr
(Rose \& Agostinho 1993; Marques \& Schuster 1994;
Gratton et al.; Pettinger et al.).
The estimated ages for our group of five stars are, however, not very
high: 2, 6, 6, 7 and 11\,Gyr.
Ages derived for giants are quite sensitive to systematic errors,
as stressed in Sect.\,\ref{uncertainage} but also the subgiant, HD\,18907,
has an estimated age of only 6\,Gyr.

\section{Conclusions}
\label{conclude}

\subsection{The use of subgiants for studies of galactic evolution}

A main result of the present study is that subgiant stars may be
straightforwardly used as probes of the evolution of many chemical elements in
the galactic disk.
This is seen from the generally small scatter among the solid dots in
Figs.\,\ref{allab}, \ref{oxfig} and \ref{alphafeteff}a which are quantified
in Table\,\ref{sgbscatter}.
For most investigated elements, with the notable exceptions of C and N, the
results for our subgiants fall within the scatter measured for disk dwarf stars.
We have found indications that the surface abundances of subgiants may be
affected by early dredge-up (before the canonical first dredge-up) of CN
processed material.
This should be further investigated by a larger sample of subgiant stars.
One such sample may be the sample of 121 subgiants investigated for
lithium destruction by do Nascimento et al. (2003)\nocite{nascimento}.

The usefulness of the subgiants for studies of galactic chemical evolution
is further increased by the separation of evolutionary tracks
in the HR diagram for stars of different masses, and
of isochrones for stars of different ages.
Isochrones for subgiant stars are rather ``horizontal'' in the HR diagramme,
which makes age determination quite insensitive to errors in effective
temperatures.
Stellar ages may therefore be determined from accurate absolute magnitudes
or surface gravities with only rough determinations of
effective temperature and of metallicity.

With parallaxes accurate to 5\% one may (with luminosities estimated from
apparent magnitudes) derive ages with relative accuracy better than 20\%
for these stars, provided that the heavy-element abundance, Z, is
known to 0.1\,dex. (This also assumes that the helium abundance
$Y$ is known within 0.05 and that the convection parameter
($\ell /H_{\rm p}$) is known or constant with an accuracy of about
0.5 (Gustafsson 1995)).
The age determination is not very dependent on the effective temperature as
long as the star is not too close to the turn-off point or to the giant branch,
i.e. as long as it is located on the almost horizontal part of the evolutionary
track in the HR diagramme.

Alternatively, the ages may be estimated from
the isochrones in the \Teff vs. $\log g$ diagramme, where the
surface gravities are estimated spectroscopically (e.g. from
the wings of strong spectral lines) or photometrically (e.g.
from the Str\"omgren $c_1$ index). With an accuracy of $\log g$
of about 0.1\,dex, which is difficult but obtainable today, one may then
estimate ages with an accuracy of about 20\% (cf. Gustafsson 1999).

So, obviously the subgiants on the horizontal part of the evolutionary
track offer possibilities comparable to or even better than those of the dwarfs
close to the turn-off point, exploited in their study of 189 stars by
Edvardsson et al. (1993).

An important advantage is also that the subgiants are typically
1-2 magnitudes brighter than the dwarfs, so that more distant galactic regions
may be explored at a similar observational cost.

The chemically distinct group of five stars which show abundances like
those of thick-disk stars, but for which we derive unexpectedly low ages,
deserves a closer study of a larger sample.
Is there a genuinely ``young'' group of $\alpha$-element rich stars,
or will a larger or more detailed  study show that this is just an effect of
small-sample statistics and uncertainties in the evolutionary tracks of giant
stars?

How many suitable stars are possible to find?
By cross-correlating the {\sc Hipparcos} catalogue
with the $uvby\beta$ catalogue by Hauck \& Mermilliod (1998)
we find about $5 \cdot 10^3$ stars in the interval $2.54<\beta<2.79$,
and with absolute magnitudes in the interval 4.3--2.0\,mag, and an
accuracy in distance better than 20\%.
The limits were set to avoid both turn-off point stars and stars on
the giant branch, and the fraction of reddened main-sequence stars
is judged to be minor. These stars are then generally within a distance
of 250\,pc, are brighter than 11.5\,mag apparently and may
thus be reached at high spectral resolution at telescopes of the
4-meter class. One may estimate that about 10-20 of these
stars are metal poor and belong to the Halo population. Obviously,
these stars together form an excellent sample for studying the
the halo -- thick disk -- thin disk transitions in the Galaxy.
In order to optimize samples for these studies rough metallicities
are needed, as well as reddening data.
These, as well as colour indices
for estimating \Teff and $\log g$, may be found from $uvby\beta$ photometry
In addition to these data, kinematic data is needed.
The {\sc Hipparcos} catalogue gives proper motions and radial velocities are
available (Barbier-Brossat \& Figon 2000, Malaroda et al. 2001).

The subgiants may also be exploited in studies
of the chemical evolution and in particular the thick-disk thin-disk
transition, using high spectral resolution for stars at distances of up to
about 1\,kpc, e.g. in the Galactic Pole regions, with the VLT and other large
telescopes.
We estimate that one should be able to find several tens of subgiants per
square degree with $V \approx 18$\,mag at a distance of $1 \pm 0.1$\,kpc
from the Galactic Plane.
As a preparatory to such studies one must, however,
compose finding lists for survey work, e.g. using $uvby$ photometry.
Here, the main objective will be to separate the subgiants from dwarfs and
red horizontal branch stars of similar colours.
The main criterion, in addition to a colour such as $b-y$ to define the
temperature interval, will be a surface gravity or luminosity criterion such
as the Balmer-discontinuity-index $c_1$.
For the task to separate the subgiants, an accuracy of $0.5$\,dex is needed in
$\log g$, corresponding (at 5500\,K) to about 0.03\,mag in $c_1$ which is not
very difficult.
For this to work, however, a relatively good metallicity
indicator must also be measured, such as the $m_1$ index.
Alternatively, the $f$ index of the narrow-band Brorfelde $gnkmf$ system could
be used for estimating the gravity
(cf. Gustafsson \& Bell 1979, and references given therein).

A third possibility is to exploit the subgiants in studies at lower
spectral resolution.
This possibility should be explored further, e.g. for
studies towards the centre of the Galaxy.

\begin{acknowledgement}
This project was supported by the Swedish Natural Science Research Council, NFR.
R. D\"ummler and I. Ilyin are thanked for operating the NOT telescope
and SOFIN software during the La Palma observation run and
for taking additional solar spectra.
J. Setiawan and S. Udry are thanked for supplying unpublished data concerning the
kinematics of HD\,18907 and HD\,62644.
Thanks are due to John Beckman for valuable comments to the manuscript, and to an
anonymous referee for suggesting several improvements.

We have made use of data from the {\sc Hipparcos} mission, SIMBAD, the NASA ADS
and the VALD database.
\end{acknowledgement}

\end{document}